\documentclass[runningheads,orivec]{llncs}
\pdfpagesattr{/CropBox [92 92 523 748]}

\usepackage{generic}
\usepackage{macros}
\iftrue 
  \newenvironment{journal}{\ignorespaces}{\ignorespacesafterend}
  \newenvironment{conference}{\comment}{\endcomment\ignorespacesafterend}
  \newenvironment{journalitemize}{\begin{itemize}\ignorespaces}{\end{itemize}}
  \newcommand{\ifjournal}[1]{#1}
  \newcommand{\ifjournalelse}[2]{#1}
  \newcommand{\ifconference}[1]{}

\else
  \newenvironment{journal}{\comment}{\endcomment\ignorespacesafterend}
  \newenvironment{conference}{\ignorespaces}{\ignorespacesafterend}
  \newenvironment{journalitemize}{\begin{enumerate*}\ignorespaces}{\end{enumerate*}}
  \newcommand{\ifjournal}[1]{}
  \newcommand{\ifjournalelse}[2]{#2}
  \newcommand{\ifconference}[1]{#1}

  \makeatletter
  \renewcommand{\xrightarrow}[2][]{\smash{\@transition\rightarrowfill[#1]{#2}}}
  \makeatother
  \renewenvironment{proof}{\comment}{\endcomment\ignorespacesafterend}
  \renewenvironment{proofsketch}{\comment}{\endcomment\ignorespacesafterend}
  \renewenvironment{compactproof}{\comment}{\endcomment\ignorespacesafterend}
  
  \renewcommand{\footnote}[1]{}
  \renewcommand{\subsection}[1]{}

  \usepackage{xpatch}
  \xapptocmd\normalsize{%
    \abovedisplayskip=6pt plus 2pt minus 4pt
    \belowdisplayskip=6pt plus 2pt minus 4pt
  }{}{}
\fi

\title{Between Linearizability and Quiescent Consistency%
  \thanks{Research supported by NSF 0916741.  \ifconference{\protect\\ The full version of this
    paper is available at \url{http://arxiv.org/abs/1402.4043}.}}}
\subtitle{Quantitative Quiescent Consistency}
\author{Radha Jagadeesan \and James Riely}
\institute{DePaul University}

\begin{document} 
\hidepolarity
\maketitle
\begin{abstract}

Linearizability is the de facto correctness criterion for concurrent
data structures.  Unfortunately, linearizability imposes a performance
penalty which scales linearly in the number of contending threads.
Quiescent consistency is an alternative criterion which guarantees
that a concurrent data structure behaves correctly when accessed
sequentially. Yet quiescent consistency says very little about
executions that have any contention.

\QQUAD We define quantitative quiescent consistency (QQC), a
relaxation of linearizability where the degree of relaxation is
proportional to the degree of contention.  When quiescent, no relaxation
is allowed, and therefore QQC refines quiescent consistency, unlike
other proposed relaxations of linearizability. We show that high
performance counters and stacks designed to satisfy quiescent
consistency continue to satisfy QQC.  The precise assumptions under
which QQC holds provides fresh insight on these structures. To
demonstrate the robustness of QQC, we provide three natural
characterizations and prove compositionality.

\end{abstract}

\section{Introduction}

This paper defines \emph{Quantitative Quiescent Consistency (QQC)} as
a criterion that lies between
linearizability \parencite{DBLP:journals/toplas/HerlihyW90} and
quiescent consistency \parencite{DBLP:journals/jacm/AspnesHS94,DBLP:journals/tocs/ShavitZ96,HS08}.
The following example should give some intuition about these criteria.

\begin{example}
\label{ex:intro}
Consider a counter object with a single \texttt{getAndIncrement} method.  The
counter's sequential behavior can be defined as a set of strings such
as
\begin{math}
  {\YINC1?0}  
  {\YINC3?1}
  {\BINC2?2}
  {\EINC2!2}
\end{math}
where $\BINC1??$ denotes an invocation (or call) of the method and
$\EINC1!i$ denotes the response (or return) with value $i$.  Suppose
each invocation is initiated by a different thread.

A concurrent execution may have overlapping method invocations.  For
example, in 
\begin{math}
  {\BINC2?2}
  {\YINC1?0}  
  {\YINC3?1}
  {\EINC2!2}
\end{math}
the execution of 
${\YINC2?2}$ 
overlaps with both
${\YINC1?0}$ 
and
${\YINC3?1}$, 
whereas 
${\YINC1?0}$ 
finishes executing before
${\YINC3?1}$
begins.
Consider the following four executions.
\begin{align*}
  {\BINC2?2}
  {\YINC1?0}  
  {\YINC3?1}
  {\EINC2!2}
  &&
  {\BINC2?2}
  {\YINC3?1}
  {\YINC1?0}  
  {\EINC2!2}
  &&
  {\BINC1?0}
  {\YINC2?2}  
  {\YINC3?1}
  {\EINC1!0}
  &&
  {\BINC1?0}
  {\YINC2?2}  
  {\EINC1!0}
  {\YINC3?1}
\end{align*}

\emph{Linearizability}
states roughly that \emph{every} response-to-invocation order in a
concurrent execution must be consistent with the sequential
specification.  Thus, the first execution
is linearizable, since the response of ${\YINC1!0}$ precedes the
invocation of ${\YINC3?1}$ in the specification.  However, none of the
other executions is linearizable.  For example, the response of ${\YINC3!1}$ precedes
the invocation of ${\YINC1?0}$ in the second execution\ifjournal{,
  but not in the specification}.

Linearizability can also be understood in terms the
\emph{linearization point} of a method execution, which must occur
between the invocation and response.  From this perspective, the first
execution above is linearizable because we can find a sequence of
linearization points that agrees with the specification; this requires
only that the linearization point of ${\YINC2?2}$ follow that of
${\YINC3?1}$.  No such sequence of linearization points exists for the
two other executions.

\emph{Quiescent consistency} is similar to linearizability, except
that the response-to-in\-vo\-ca\-tion order must be respected only
across a quiescent point, that is, a point with no open method calls.
The first three executions above are quiescently consistent \ifjournal{simply}
because there are no non-trivial quiescent points.  The last execution
fails to be quiescently consistent since the order from ${\YINC2!2}$
to ${\YINC3?1}$ is not preserved in the specification.

We define \emph{Quantitative Quiescent Consistency (QQC)} to require
that the number of response-to-invocation pairs that are out-of-order
at any point be bounded by the number of open calls that might be
ordered later in the specification.  We
also give a \emph{counting characterization} of QQC, which requires
that if a response matches the $i^{\textit{th}}$ method call in the
specification, then it must be preceded by at least $i$ invocations.

The first two executions above are QQC; however, the last two are not.  In
the second execution,
the open call to ${\YINC2?2}$ justifies the return of
${\YINC3?1}$ before ${\YINC1?0}$ since ${\YINC2?2}$ occurs after
${\YINC3?1}$ in the specification.  However, in the third execution,
the return of ${\YINC2?2}$ before ${\YINC3?1}$ cannot be justified
only by the call to ${\YINC1?0}$ since ${\YINC1?0}$ occurs earlier in
the specification.
Following the counting characterization sketched above, the third
execution fails since ${\YINC2?2}$ is the third method call in the
specification trace, but the response of ${\YINC2!2}$ is only
preceded by two invocations: ${\BINC1?0}$
and ${\BINC2?2}$.
\end{example}

Quiescent consistency is too coarse to be of much use in reasoning
about concurrent executions.  For example, a sequence of interlocking
calls never reaches a quiescent point; therefore it is trivially
quiescently consistent.  This includes obviously correct executions, such as
\begin{math}
  {\BINC1??}
  {\BINC2??}
  {\EINC1!0} 
  {\BINC1??}
  {\EINC2!1} 
  {\BINC2??}
  {\EINC1!2} 
  {\BINC1??}
  {\EINC2!3} 
  {\BINC2??}
  {\EINC1!4} 
  {\BINC1??}
  \cdots,
\end{math}
nearly correct executions, such as
\begin{math}
  {\BINC1??}
  {\BINC2??}
  {\EINC1!1} 
  {\BINC1??}
  {\EINC2!0} 
  {\BINC2??}
  {\EINC1!3} 
  {\BINC1??}
  {\EINC2!2} 
  {\BINC2??}
  {\EINC1!5} 
  {\BINC1??}
  \cdots,
\end{math}
and also ridiculous executions, such as 
\begin{math}
  \BINC1??
  \BINC2??
  \EINC1!{1074}
  \BINC1??
  \EINC2!{17}
  \BINC2??
  \EINC1!{2344}
  \BINC1??
  \EINC2!{3}
  \BINC2??
  \cdots.
\end{math}

Linearizability has proven quite useful in reasoning about concurrent
executions; however, it fundamentally constrains efficiency in a
multicore setting: \textcite{DBLP:journals/jacm/DworkHW97} show that
if many threads concurrently access a linearizable counter, there must
be either a location with high contention or an execution path that
accesses many shared variables.
\ifjournal{\par}
\textcite{Shavit:2011:DSM:1897852.1897873} argues that the performance
penalty of linearizable data structures is increasingly unacceptable
in the multicore age.  This observation has lead to a recent renewal
of interest in nonlinearizable data structures.  As a simple example,
consider the following counter implementation: a simplified version of
the counting networks of \textcite{DBLP:journals/jacm/AspnesHS94}.

\begin{lstlisting}[numbers=none]
          class #\ctype{Counter<N:Int>}# {
             field b:#\ctype{[0..N-1]}# = 0;                // 1 balancer
             field c:#\ctype{Int[]}#    = [0, 1, ..., N-1]; // N counters
             method getAndIncrement():Int {
                val i:#\ctype{[0..N-1]}#; 
                atomic { i = b; b++; }
                atomic { val v = c[i]; c[i] += N; return v; } } }
\end{lstlisting}
The \texttt{$N$-Counter} has two fields: a \emph{balancer} \texttt{b}
and an array \texttt{c} of $N$ integer counters.  There are two atomic
actions in the code: The first reads and updates the balancer, setting
the local index variable \texttt{i}.  The second reads and updates the
\texttt{i}$^{\textit{th}}$ counter.  Although the balancer has high
contention in our simplified implementation, the counters do not;
balancers that avoid high contention are described
in \parencite{DBLP:journals/jacm/AspnesHS94}.

\begin{example}
\label{ex:count1}
The \texttt{$N$-Counter} behaves like a sequential counter if calls to
\texttt{get\-And\-Inc\-re\-ment} are sequentialized.  To see this, consider a
\texttt{$2$-Counter}, with initial state $\cstate{0}{0}{1}$.  In a
series of sequential calls, the state progresses as follows, where we
show the execution of the first atomic with the invocation and the
second atomic with the response.  The execution 
\begin{math}
  \YINC1?0 
  \YINC3?1
  \BINC2?2
  \EINC2!2
\end{math}
can be elaborated as follows.
\begin{alignat*}{3}
  \cstate{0}{0}{1}
    &\xrightarrow{\BINC1?0} &\cstate{1}{0}{1} &\xrightarrow{\EINC1!0} &\cstate{1}{2}{1}
  \\&\xrightarrow{\BINC3?1} &\cstate{0}{2}{1} &\xrightarrow{\EINC3!1} &\cstate{0}{2}{3}
  \\&\xrightarrow{\BINC2?2} &\cstate{1}{2}{3} &\xrightarrow{\EINC2!2} &\cstate{1}{4}{3}
\end{alignat*}
When there is concurrent
access,
the \texttt{$2$-Counter} allows nonlinearizable executions, such as 
\begin{math}
  \BINC2?2
  \YINC3?1
  \YINC1?0  
  {\EINC2!2.}
\end{math}
\begin{alignat*}{3}
  \cstate{0}{0}{1}
    &\xrightarrow{\BINC2?2} &\cstate{1}{0}{1} 
  \\&\xrightarrow{\BINC3?1} &\cstate{0}{0}{1} &\xrightarrow{\EINC3!1} &\cstate{0}{0}{3}
  \\&\xrightarrow{\BINC1?0} &\cstate{1}{0}{3} &\xrightarrow{\EINC1!0} &\cstate{1}{2}{3}
  \\&&&\xrightarrow{\EINC2!2} &\cstate{1}{4}{3}
\end{alignat*}
With a sequence of interlocking calls, it is also possible for the
\texttt{$N$-Counter} to execute as
\begin{math}
  {\BINC1??}
  {\BINC2??}
  {\EINC1!1} 
  {\BINC1??}
  {\EINC2!0} 
  {\BINC2??}
  {\EINC1!3} 
  {\BINC1??}
  {\EINC2!2} 
  {\BINC2??}
  {\EINC1!5} 
  {\BINC1??}
  \cdots,
\end{math}
producing an infinite sequence of values that are just slightly out of
order.  Using the results of this paper, one can conclude that with a
maximum of two open calls, the value returned by
\texttt{getAndIncre\-ment} will be ``off'' by no more than 2, but this
does not follow from quiescent consistency.
\end{example}

Our results are related to those
of \parencite{DBLP:journals/jacm/AspnesHS94,DBLP:journals/mst/ShavitT97,DBLP:journals/cjtcs/AielloBHMST00,DBLP:conf/podc/BuschM96}.
In particular, \textcite{DBLP:journals/jacm/AspnesHS94} prove that in
any \emph{quiescent} state (with no call that has not returned), such
a counter has a ``step-property'', indicating the shape of \texttt{c}.
Between $\EINC3!1$ and $\EINC1!0$ in the second displayed execution of
\autoref{ex:count1}, the states with $\cystate{0}{0}{3}$ do \emph{not} have
the step property, since the two adjacent counters differ by
more than $1$.

\citeauthor{DBLP:journals/jacm/AspnesHS94} imply that the step
property is related to quiescent consistency\ifjournalelse{, but they do not formally
state this.  Indeed, they do not provide a formal definition of
quiescent consistency.}{, but they do not provide a formal definition.}  It appears that they have in mind
is something like the following: An execution is \emph{weakly
  quiescent consistent} if any uninterrupted subsequence of \emph{sequential} calls
(single calls separated by quiescent points) is a subtrace of a
specification trace.

The situation is delicate: Although the increment-only counters of
\parencite{DBLP:journals/jacm/AspnesHS94} are quiescently consistent
in the sense we defined in \autoref{ex:intro} (indeed, they
are QQC), the increment-decrement counters of
\parencite{DBLP:journals/mst/ShavitT97,DBLP:journals/cjtcs/AielloBHMST00,DBLP:conf/podc/BuschM96}
are only \emph{weakly} quiescent consistent.  Indeed, the theorems proven
in \parencite{DBLP:journals/mst/ShavitT97}
state only that, at a quiescent point, a variant of the step property
holds.  They state nothing about the actual values read from the
individual counters.  Instead, our definition requires that a quiescently consistent execution be
a permutation of \emph{some} specification trace, even if it has no
nontrivial quiescent points.
\begin{example}
  \label{ex:count2}
  Consider an extension of the
  \texttt{$2$-Counter} with \texttt{decrementAndGet}.
\begin{lstlisting}[numbers=none]
                method decrementAndGet():Int {
                   val i:#\ctype{[0..N-1]}#; 
                   atomic { i = b-1; b--; }
                   atomic { c[i] -= N; return c[i]; } }
\end{lstlisting}
The execution 
\begin{math}
{\BINC1??}
{\BINC3??}
{\BDEC2??}
{\BDEC4??}
{\EDEC4!{-2}}
{\EINC1!{-2}}
{\EINC3!1}
{\EDEC2!1}
\end{math}
is possible, although this is not a permutation of any specification
trace.  The execution proceeds as follows.
\begin{alignat*}{3}
  \cstate{0}{0}{1}
    &  \xrightarrow{\BINC1??} &\cxstate{1}{0}{1} 
    &  \xrightarrow{\BINC3??} &\cstate{0}{0}{1} 
  \\&  \xrightarrow{\BDEC2??} &\cxstate{1}{0}{1} 
    &  \xrightarrow{\BDEC4??} &\cstate{0}{0}{1} 
  \\&  \xrightarrow{\EDEC4!{-2}} &\cxstate{0}{-2}{1}
    &  \xrightarrow{\EINC1!{-2}} &\cstate{0}{0}{1}
  \\&  \xrightarrow{\EINC3!1} &\cxstate{0}{0}{3}
    &  \xrightarrow{\EDEC2!1} &\cstate{0}{0}{1}
  \amsqed
\end{alignat*}
\end{example}
It is important to emphasize that this increment-decrement counter is
not even quiescently consistent according to our definition.  There is no hope that it could
satisfy any stronger criterion.

Of course counters are not the only data structures of interest.   In
\ifjournalelse{this paper}{the full paper}, we treat concurrent stacks in detail.
We define a simplified \texttt{$N$-Stack} below; the full, tree-based data
structure is defined in
\textcite{DBLP:journals/mst/ShavitT97}\ifjournal{ and summarized in \autoref{sec:stack}}. 
\begin{lstlisting}[numbers=none]
        class #\ctype{Stack<N:Int>}# {
           field b:#\ctype{[0..N-1]}# = 0;                 // 1 balancer
           field s:#\ctype{Stack[]}#  = [[], [], ..., []]; // N stacks of values
           method push(x:Object):Unit {
              val i:#\ctype{[0..N-1]}#;
              atomic { i = b; b++; }
              atomic { val v = s[i].push(x); return v; } }
           method pop():Object {
              val i:#\ctype{[0..N-1]}#;
              atomic { i = b-1; b--; }
              atomic { val v = s[i].pop(); return v; } } }
\end{lstlisting}
The trace given in \autoref{ex:count2} for the increment-decrement
counter is also a trace of the stack, where we interpret \texttt{+} as
\texttt{push} and \texttt{-} as \texttt{pop}.
Whereas this is a nonsense execution for a
counter, it is a linearizable execution of a stack: simply choose the
linearization points so that each push occurs immediately before the corresponding pop.  Nonetheless, the
\texttt{$N$-Stack} is only \emph{weakly} quiescent consistent in
general.

\begin{example}
  \label{ex:2stack:notqqc1}
  The \texttt{$N$-Stack} generates the execution
  \begin{math}
    \YPUT1.a
    \YPUT2.b
    \BPUT3?c
    \YGET4.a
    \EPUT3!c
  \end{math}
  as follows.
\begin{alignat*}{3}
  \szstate{0}{}{}
    &  \xrightarrow{\BPUT1?a} &\szstate{1}{}{} 
    &  \xrightarrow{\EPUT1!a} &\szstate{1}{\COLOR1a}{} 
  \\&  \xrightarrow{\BPUT2?b} &\szstate{0}{\COLOR1a}{} 
    &  \xrightarrow{\EPUT2!b} &\szstate{0}{\COLOR1a}{\COLOR2b} 
  \\&  \xrightarrow{\BPUT3?c} &\szstate{1}{\COLOR1a}{\COLOR2b}
  \\&  \xrightarrow{\BGET4?a} &\szstate{0}{\COLOR1a}{\COLOR2b}
    &  \xrightarrow{\EGET4!a} &\szstate{0}{}{\COLOR2b}
  \\&&&\xrightarrow{\EPUT3!c} &\szstate{0}{\COLOR3c}{\COLOR2b}
\end{alignat*}
  However, this specification is not quiescently consistent with any stack
  execution:  There is a quiescent point
  after each of the first two pushes;  therefore it is impossible to pop
  $\COLOR1a$ before $\COLOR2b$.  This execution is possible even when
  there are several pushes beforehand.
\end{example}
  
In the case of the \texttt{$N$-Stack}, a simple \emph{local}
constraint can be imposed in order to establish quiescent consistency:
intuitively, we require that no pop \emph{overtakes} a push on the
same stack \texttt{s[i]}.  In 
\ifjournalelse{\autoref{sec:stack}}{the full paper,} 
we show that the
stack is actually \emph{QQC} under this constraint, and therefore
quiescently consistent.  
We also prove that the elimination-tree stacks of
\textcite{DBLP:journals/mst/ShavitT97} are QQC.
The increment-only
counters of \parencite{DBLP:journals/jacm/AspnesHS94} are also QQC\ifjournalelse{, although in this
case, we have elided the proofs: The}{; the} proofs for the tree-based
increment-only counter follow the structure of the proofs for the
elimination-tree stacks.
(We have not found a \emph{local} constraint
under which the increment-decrement counter is quiescently consistent\ifjournal{;
we believe that it may be achievable with a global toggle that
determines how to resolve the races at each point, but this, of
course, defeats the point}.)
Our correctness result is much stronger than that of
\parencite{DBLP:journals/mst/ShavitT97}, which only proves \emph{weak}
quiescent consistency.  
\ifjournal{\bigskip}

The preliminary version of \citeauthor{DBLP:journals/mst/ShavitT97}'s
paper \parencite{DBLP:conf/spaa/ShavitT95} suggests an upcoming
definition \emph{$\epsilon$-linearizability}, 
``a variant of linearizability that
  captures the notion of `almostness' by allowing a certain fraction
  of concurrent operations to be out-of-order.''
\ifjournal{Since the details 
did not make it into the final version of the paper
\parencite{DBLP:journals/mst/ShavitT97}, it is unclear whether the
``fraction of concurrent operations'' is meant to vary depending on the amount
of concurrency available at any given moment, or if the
``fraction'' is fixed at the outset.  If it is meant to vary, then it
would be very similar to QQC.
\par}
This thread was picked up by \textcite{DBLP:conf/opodis/AfekKY10} and
improved by \textcite{DBLP:conf/popl/HenzingerKPSS13}.  As defined in
\parencite{DBLP:conf/popl/HenzingerKPSS13}, the idea is to define a cost
metric on relaxations of strings and to bound the relaxation cost for
the specification trace that matches an execution.  This
relaxation-based approach has been
used to validate several novel concurrent data
structures \parencite{DBLP:conf/opodis/AfekKY10,DBLP:conf/cf/HaasLHPSKS13}.
With the exception of the
increment-only counter validated in
\parencite{DBLP:conf/opodis/AfekKY10}, all of these data
structures intentionally violate quiescent consistency.
In \ifjournalelse{\autoref{sec:qqc:compare}}{\autoref{sec:qqc}}, we show
that this approach in incomparable to QQC.

With QQC, the maximal degradation depends
upon the amount of concurrent access, whereas in the relaxation-based
approach it does not.  Thus, QQC ``degrades gracefully'' as
concurrency increases.  In particular, a QQC data structure that is
accessed sequentially will exactly obey the sequential specification,
whereas a data structure validated against the relaxation-based
approach may not.

\ifjournal{\bigskip}

In the rest of the paper, we formalize QQC and study its properties.
\ifjournal{The heart of the paper is \autoref{sec:qqc}, which defines QQC and
establishes its properties.  
The impatient reader can safely skim up to
that section, referring back as necessary.\par}
Our contributions are as follows.
\ifjournalelse{\begin{itemize}}{\begin{itemize}[nosep]}
\item We define linearizability (\autoref{sec:linear}), quiescent
  consistency (\autoref{sec:qc}) and QQC (\autoref{sec:qqc}) in terms
  of partial orders over events with duration.  
  \ifjournalelse{The formalities of the model are described in \autoref{sec:model}.}
  As in \autoref{ex:intro}, the definitions are given in
  terms of the order from response to invocation.

\item For sequential specifications, we provide alternative characterizations of linearizability, quiescent
  consistency and QQC in terms of
  the number of invocations that precede a response.  
  \ifjournal{This is the characterization used in most proofs.}
  For linearizability, this approach can be found in \parencite{BDG13}.

\item We provide an alternative characterization of QQC in terms of a
  proxy that controls access to the underlying sequential data
  structure.  The proxy adds a form of \emph{speculation} to the flat
  combining technique of \textcite{DBLP:conf/spaa/HendlerIST10}.  This
  characterization can be seen as a language generator, rather than an
  accepter.  \ifjournal{We show that the proxy is sound and complete
    for QQC; that is, it generates {\em all and only} traces that are
    QQC.}

\item Like linearizability and quiescent consistency \parencite{HS08},
  QQC is non-blocking and compositional. Like quiescent consistency
  and unlike linearizability, a QQC execution may not respect program
  order, and therefore QQC is incomparable to sequential consistency \cite{Lamport}.
  We prove that QQC is compositional for sequential specifications, in the sense of \textcite{DBLP:journals/toplas/HerlihyW90}.

\item We show that QQC is useful for reasoning about data structures
  in the literature.  In 
  \ifjournalelse{\autoref{sec:stack}}{the full paper},
  we prove that the elimination tree stacks of
  \textcite{DBLP:journals/mst/ShavitT97} are QQC, as long as no pop
  overtakes a push on the same stack. 
\end{itemize}


\begin{journal}
\section{Model}
\label{sec:model}

The semantics of a concurrent program is given as a process.  A
\emph{process} is a set of traces.  A \emph{trace} is a finite,
polarized LPO (labelled partial order).  Formally, we define traces to
be finite sets of named \emph{events}.  The event names are the
carrier set for the LPO, and the order is embedded in the events
themselves using name sets.

\subsection{Events}
\label{sec:events}

An event is a quadruple, consisting of a polarity, a label, a name
(identifying a node the partial order) and a set of names (identifying
the preceding nodes in the partial order).  As a standard example, the
reader may want to consider labels generated by the grammar
\begin{math}
  \aprim
  \BNFDEF
  \primcallone{\athrd}{\aobj\,\afun}{\aval}
  \BNFSEP
  \primretone{\athrd}{\aobj\,\afun}{\aval}
\end{math}
where $\athrd$ is a thread identifier,
$\aobj$ is an object name,
$\afun$ is a function name, and
$\aval$ is the actual parameter or return value.

Let $\aact\CC\bact\in\Act$ range over names and
$\aactset\CC\bactset\subseteq\Act$ range over finite sets of names.  And
let $\aprim\in\Prim$ range over labels (with some interpretation in
the application domain).  Then events are defined as follows\footnote{In
  this paper, we consider the simple case of non-interacting
  composition.  This allows us to ignore the internal polarity which
  arise from the interaction of input and output.}.
\begin{align*}
  \albl\CC\blbl
  \BNFDEF\eventtuple?{\aprim}{\aact}{\aactset}
  \BNFSEP\eventtuple{\bact}{\aprim}{\aact}{\aactset}
\end{align*}
Under our standard example, we would expect events to come in pairs of
the form
$\eventtuple?{\primcallone{\athrd}{\aobj\,\afun}{\aval}}{\aact}{\aactset}$
and
$\eventtuple{\aact\,}{\primretone{\athrd}{\aobj\,\afun}{\bval}}{\bact}{\bactset}$,
where $\aval$ is the actual parameter and $\bval$ is the returned
value.  The appearance of $\aact$ in the return event indicates that
this event closes the open call named $\aact$.

Three of the components in an event can be retrieved simply.  We use
the following functions:
$\feprim{\eventtuple{\pinp}{\aprim}{\aact}{\aactset}}\eqdef\aprim$,
$\feid{\eventtuple{\pinp}{\aprim}{\aact}{\aactset}}\eqdef\aact$ and
$\febefore{\eventtuple{\pinp}{\aprim}{\aact}{\aactset}}\eqdef\aactset$.
For the remaining component, we define both the functions $\fepol{}$ and
$\febrak{}$.  Let $\apol\in\set{\pinp\CC\pout}$
range over the polarities for input (\pinp) and output (\pout) and let
$\actnone$ be a reserved name.
\begin{align*}
  \fepol{\albl}
  \eqdef
  \begin{cases}
    \pinp &\text{if } \albl = \eventtuple{\pinp}{\aprim}{\aact}{\aactset}
    \\
    \pout &\text{if } \albl = \eventtuple{\bact}{\aprim}{\aact}{\aactset}
  \end{cases}
  &&
  \febrak{\albl}
  \eqdef
  \begin{cases}
    \actnone &\text{if } \albl = \eventtuple{\pinp}{\aprim}{\aact}{\aactset}
    \\
    \bact &\text{if } \albl = \eventtuple{\bact}{\aprim}{\aact}{\aactset}
  \end{cases}
\end{align*}
Because the standard example is so familiar, we will consider
invocation/call/in\-put/\pinp{} to be synonymous, and likewise
response/return/output/\pout{}.

We sometimes use superscripts on name metavariables, such as
$\xact!{}$ and $\xact?{}$.  Any name bound to $\xact!{}$ must be
associated with an output event, and likewise for input events.  The
superscript makes these distinct metavariables. Thus we have
$\xact!{}\neq\xact?{}$.

Turning to the order between events, we write $\albl\lthb\blbl$ to
indicate that $\albl$ precedes $\blbl$:
\begin{math}
  (\albl\lthb\blbl)
  \eqdef
  \feid\albl\in\febefore\blbl.
\end{math}

\subsection{Traces }
\label{sec:traces}

We use $\ptr$--$\btr$ to range over \emph{event sets} (finite
sets of events).  
Define
\begin{math}
  \feids{\atr}\eqdef
  \setst{\feid\albl}{{\albl\in\atr}}
\end{math}
and let $\aact\in\atr$ be shorthand for $\aact\in\feids\atr$.

Given an event set $\atr$ and name set $\aactset$, define \emph{indexing} as
\begin{math}
  \feindex\atr\aactset\eqdef
  \setst{\albl\in\atr}{\feid\albl\in\aactset}.
\end{math}
Thus
\begin{math}
  \feindex\atr{\feids{\atr}}=\atr.
\end{math}
If event names are unique, this generates the partial function
$\feindex\atr\aact$ for single names: if
$\feindex\atr{\set\aact}=\emptyset$ then $\feindex\atr\aact$ is
undefined; if $\feindex\atr{\set\aact}=\set{\albl}$ then
\begin{math}
  \feindex\atr\aact\eqdef\albl.
\end{math}
Indexing provides a natural way to lift ordering relations from events
to names:
\begin{math}
  (\aact\lttrans\atr\bact) \eqdef
  (\feindex\atr\aact\lthb\feindex\atr\bact).
\end{math}
Let be $\leqtrans{\atr}$ the reflexive closure of $\lttrans{\atr}$.

An event set $\atr$ is 
a \emph{trace} 
if it satisfies the following, $\forall\albl\CC\blbl\in\atr$.
\begin{enumerate}
\item event names are unique: if $\feid\albl=\feid\blbl$ then
  $\albl=\blbl$
\item\label{trace:before} $\febefore{}$ okay:
  $\forall\aact\in\febefore{\albl}$.
  $\exists\clbl\in\atr$.
  $\aact=\feid{\clbl}$
\item\label{trace:brak} $\febrak{}$ okay:
  if $\fepol{\albl}=\pout$ then
  $\febrak{\albl}\in\febefore{\albl}$ and
  $\fepol{\feindex\atr{\febrak{\albl}}}=\pinp$
\item\label{trace:inp} input acquires control: if
  $\aact\lttrans{\atr}\zact?2$ then
  $\exists\zact!3$.
  $\aact\leqtrans{\atr}\zact!3\lttrans{\atr}\zact?2$
\item\label{trace:out} output releases control: if
  $\zact!1\lttrans{\atr}\bact$ then
  $\exists\zact?3$.
  $\zact!1\leqtrans{\atr}\zact!3\lttrans{\atr}\bact$
\item $\lttrans{\atr}$ defines a strict partial order
  (irreflexive, antisymmetric and transitive)
\end{enumerate}

A trace $\atr$ is \emph{\operational{}} if 
$\forall\zact?1\CC\zact!2\in\atr$.  either
$\zact?1\lttrans{\atr}\zact!2$ or $\zact!2\lttrans{\atr}\zact?1$.

A trace $\atr$ is \emph{sequential} if
$\forall\aact\CC\bact\in\atr$. either $\aact\lttrans{\atr}\bact$ or $\bact\lttrans{\atr}\aact$.

\bigskip

Our model can be viewed as a labelled partial order enriched with polarity and
bracketing.  Most significant here are requirements \eqref{trace:inp}
and \eqref{trace:out} in the definition of a trace.  One immediate
consequence is that input events cannot be related to other input
events unless there is an intervening output event, and similarly for
the dual case.  

Consider two bracketed event sequences $\B1?\E1!$ and $\B2?\E2!$.  As
indicated by condition \eqref{trace:brak} in the definition of traces,
the open brackets must be \pinp{} events.  There are six possible
relations among the events.  Three of these are familiar: it could be
that $\B1?\E1!$ precedes $\B2?\E2!$, or that $\B2?\E2!$ precedes
$\B1?\E1!$ or that they are concurrent.
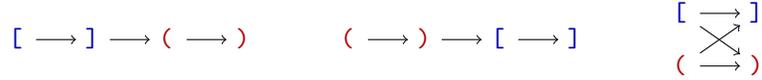
\begin{align*}
  \begin{tikzpicture}[xscale=1,yscale=.7]
    \node (b1) at (0,0) {\B1?};
    \node (e1) at (1,0) {\E1!};
    \node (b2) at (2,0) {\B2?};
    \node (e2) at (3,0) {\E2!};
    \draw [query] (b1) to (e1);
    \draw [query] (b2) to (e2);
    \draw [bang] (e1) to (b2);
  \end{tikzpicture}
  &&
  \begin{tikzpicture}[xscale=1,yscale=.7]
    \node (b1) at (0,0) {\B2?};
    \node (e1) at (1,0) {\E2!};
    \node (b2) at (2,0) {\B1?};
    \node (e2) at (3,0) {\E1!};
    \draw [query] (b1) to (e1);
    \draw [query] (b2) to (e2);
    \draw [bang] (e1) to (b2);
  \end{tikzpicture}
  &&
  \begin{tikzpicture}[xscale=1,yscale=.7]
    \node (b1) at (0,0) {\B2?};
    \node (e1) at (1,0) {\E2!};
    \node (b2) at (0,1) {\B1?};
    \node (e2) at (1,1) {\E1!};
    \draw [query] (b1) to (e1);
    \draw [query] (b2) to (e2);
    \draw [query] (b2) to (e1);
    \draw [query] (b1) to (e2);
  \end{tikzpicture}
\end{align*}
All of these traces are fully specified in the sense that every
\pinp{} is ordered with respect to every \pout{}, and dually every
\pout{} is ordered with respect to every \pinp{}.  We call such traces
\emph{\operational{}} in that they correspond to traces generated by an
interleaving semantics.  In addition, the first two traces are
\emph{sequential}, since there is a total order on the events.  Note
that in any sequential trace, the initial event must be an input; this
follows from properties \eqref{trace:before} and \eqref{trace:brak} in
the definition of traces.

There is a homomorphism from strings of bracketed labels to
\operational{} traces: each input in the string is ordered with respect
to each output that follows it in the string, and dually.  If we
narrow attention to sequential traces, this is an isomorphism.  For
example, we can write the first two traces above as the strings
\begin{math} 
  {\B1?}
  {\E1!}
  {\B2?}
  {\E2!}  
\end{math}
and 
\begin{math} 
  {\B2?}
  {\E2!}
  {\B1?}
  {\E1!}, 
\end{math}
respectively.  The last trace above can be written as
any interleaving in
\begin{math} 
  {\B1?}
  {\E1!}
  \interleave
  {\B2?}
  {\E2!}
\end{math}
that orders the inputs before both outputs; these are
\begin{math} 
  {\B1?}
  {\B2?}
  {\E1!}
  {\E2!}, 
\end{math}
\begin{math} 
  {\B2?}
  {\B1?}
  {\E1!}
  {\E2!},  
\end{math}
\begin{math} 
  {\B1?}
  {\B2?}
  {\E2!}  
  {\E1!},
\end{math}
and
\begin{math} 
  {\B2?}
  {\B1?}
  {\E2!}  
  {\E1!}.
\end{math}
We use this notation when giving examples of \operational{} traces, as in
the introduction.

As a consequence of the homomorphism, we can use string notation on
\operational{} traces without ambiguity.  Specifically, let $\atr\,\btr$
represent the concatenation of $\atr$ and $\btr$ and
$\atr\interleave\btr$ represent the set of their interleavings,
with renaming as necessary to avoid collisions between the names of
$\atr$ and $\btr$.

Our model also allows underspecification of the relationship.
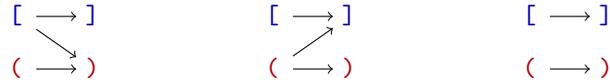
\begin{align*}
  \begin{tikzpicture}[xscale=1,yscale=.7]
    \node (b1) at (0,0) {\B2?};
    \node (e1) at (1,0) {\E2!};
    \node (b2) at (0,1) {\B1?};
    \node (e2) at (1,1) {\E1!};
    \draw [query] (b1) to (e1);
    \draw [query] (b2) to (e2);
    \draw [query] (b2) to (e1);
  \end{tikzpicture}
  &&
  \begin{tikzpicture}[xscale=1,yscale=.7]
    \node (b1) at (0,0) {\B2?};
    \node (e1) at (1,0) {\E2!};
    \node (b2) at (0,1) {\B1?};
    \node (e2) at (1,1) {\E1!};
    \draw [query] (b1) to (e1);
    \draw [query] (b2) to (e2);
    \draw [query] (b1) to (e2);
  \end{tikzpicture}
  &&
  \begin{tikzpicture}[xscale=1,yscale=.7]
    \node (b1) at (0,0) {\B2?};
    \node (e1) at (1,0) {\E2!};
    \node (b2) at (0,1) {\B1?};
    \node (e2) at (1,1) {\E1!};
    \draw [query] (b1) to (e1);
    \draw [query] (b2) to (e2);
  \end{tikzpicture}  
\end{align*}
The leftmost of these says only that $\B2?\E2!$ cannot precede
$\B1?\E1!$.  Said positively, either $\B1?\E1!$ precedes
$\B2?\E2!$, or they are concurrent.  The rightmost of these places no
constraints on the relative order of $\B1?\E1!$ and
$\B2?\E2!$.

\Operational{} traces can be seen as having a global notion of time:
everyone agrees what happened before what.  The constraints between
pairs of inputs and pairs of output simply indicate the limits of
observability: it is impossible to tell which of two calls happened
first.  In this light, one may view an underspecified trace as a
representative for the set of \operational{} traces that can be derived
by augmenting the partial order.  We take this viewpoint in our
compositionality result, which is stated only for \operational{} traces.

\bigskip

We define several notations for event sets and traces.

As noted above, for \operational{} traces $\atr$ and $\btr$ we use string notation for
concatenation ($\atr\btr$) and the set of interleavings ($\atr\interleave\btr$).

A \emph{renaming} of a trace is identical to the original trace up to
a bijection on names\footnote{$(\atr\ideq\btr)$ is defined to mean
  that there exists a bijection $\falpha{}$ on names such that
  \begin{enumerate*}
  \item $\feids{\atr}=\falpha{\feids{\btr}}$, and
  \item $\forall\aact\in\feids{\atr}$. $\feindex\atr\aact=\feindex\btr{\falpha\aact}$.
  \end{enumerate*}
  (In the first condition, we have used the obvious homomorphic
  extension of $\falpha{}$ to sets of names.)}.
We write $\ideq$ for
equivalence up to renaming.

A \emph{permutation} of a trace contains events with the same names,
labels and polarities, but may differ in ordering.  Permutation does
\emph{not} allow renaming; the names to pick out the witnessing
bijection.  We write $\permeq$ for equivalence up to permutation\footnote{%
Let 
\begin{math}
  (\albl\eqprimbrak\blbl)
  \eqdef
  (\feprim{\albl}=\feprim{\blbl})
  \land
  (\febrak{\albl}=\febrak{\blbl}).
\end{math}\\
Then define
\begin{math}
  (\atr\permeq\btr) 
  \eqdef
  (\feids{\atr}=\feids{\btr}) \land (\forall\aact\in\atr.\;\feindex\atr{\aact}\eqprimbrak\feindex\btr{\aact}).
\end{math}}.
Define $\atr\permlt\btr$ to
mean that $\atr$ is a subtrace of a permutation of $\btr$:
\begin{math}
  (\atr\permlt\btr) \eqdef
  (\exists \atr'.\;\; \atr\subseteq\atr' \permeq\btr).
\end{math}

A \emph{prefix} is a down-closed subtrace\footnote{Trace
  $\btr$ is a \emph{prefix} of trace $\atr$ if
  $\forall\aact\CC\bact\in\atr$. if ${\aact}\in{\btr}$ and
  $\bact\lttrans{\atr}\aact$ then ${\bact}\in{\btr}$.}.
We write $\btr\ltpre\atr$ or $\atr\gtpre\btr$ to indicate that
$\btr$ is a prefix of $\atr$, and
$\fedown\atr\aact$ for the smallest down-closed subset of $\atr$ that
includes $\aact$.

We treat traces both as sets of events and as partial orders.  We
use $(\setminus)$ for set difference and $(\txminus)$ for partial
order difference\footnote{An event set $\btr$ is \emph{bracketed} if every output in $\btr$ has
a matching input in $\btr$; that is
\begin{math}
  \forall\albl\in\btr.
  \textif \fepol{\albl}=\pout
  \textthen \febrak{\albl}\in\feids{\btr}.
\end{math}
A bracketed set may contain unmatched inputs, but not unmatched outputs.

For arbitrary event sets, we write $\atr\setminus\btr$ for set difference.
For trace $\atr$ and bracketed event set $\btr$, we write
$\atr\txminus\btr$ for partial order difference.  For example,
consider the trace the sequential trace
\begin{math}
  \atr=
  \COLOR1{\eventtuple{\pinp}{\aprim_1}{\aact}{\emptyset}}
  \COLOR1{\eventtuple{\aact\,}{\aprim_2}{\aact'}{\set{\aact}}}
  \COLOR2{\eventtuple{\pinp}{\aprim_3}{\bact}{\set{\COLOR1{\aact}\CC\COLOR1{\aact'}}}}
  \COLOR2{\eventtuple{\bact\,}{\aprim_4}{\bact'}{\set{\COLOR1{\aact}\CC\COLOR1{\aact'}\CC\bact}}}
  \COLOR3{\eventtuple{\pinp}{\aprim_5}{\cact}{\set{\COLOR1{\aact}\CC\COLOR1{\aact'}\CC\COLOR2{\bact}\CC\COLOR2{\bact'}}}}
\end{math}
and let $\btr$ be the bracketed set
$\set{\feindex{\atr}{\bact}\CC\feindex{\atr}{\bact'}}$.  Then we have
\begin{math}
  \atr\setminus\btr =
  \COLOR1{\eventtuple{\pinp}{\aprim_1}{\aact}{\emptyset}}
  \COLOR1{\eventtuple{\aact\,}{\aprim_2}{\aact'}{\set{\aact}}}
  \COLOR3{\eventtuple{\pinp}{\aprim_5}{\cact}{\set{\COLOR1{\aact}\CC\COLOR1{\aact'}\CC\COLOR2{\bact}\CC\COLOR2{\bact'}}}}
\end{math}
and
\begin{math}
\atr\txminus\btr =
  \COLOR1{\eventtuple{\pinp}{\aprim_1}{\aact}{\emptyset}}
  \COLOR1{\eventtuple{\aact\,}{\aprim_2}{\aact'}{\set{\aact}}}
  \COLOR3{\eventtuple{\pinp}{\aprim_5}{\cact}{\set{\COLOR1{\aact}\CC\COLOR1{\aact'}}}}.
\end{math}}.
\end{journal}

\section{Linearizability}
\label{sec:linear}

\begin{conference}
A \emph{trace} is a labelled partial order with polarity and
bracketing. We use \op? and \op! to denote polarities.  The polarity
indicates whether an event in the partial order is a call/input (\op?)
or a return/output (\op!).  Bracketing matches each return with the
particular call that precedes it.  Let $\ptr$--$\btr$ range over
traces and let $\aact\CC\bact$ range over \emph{names}, which form the
carrier set of the partial order.  We introduce notation over traces
as needed.
\end{conference}
\begin{journal}
We give two characterizations of linearizability and prove compositionality.

In \autoref{sec:lin:1}, we give a characterization that looks at every
way to \emph{cut} a trace into prefix and suffix; linearizability
requires that response-to-invocation order be respected across all
cuts.  This corresponds to characterization of QQC given in
\autoref{sec:qqc:1}.  In the case of QQC, a certain number of
invocations may be ignored, proportional to the number of calls that
are both open across the cut and out of specification-order with
respect to the response.

In \autoref{sec:lin:2} we give a subset-based characterization, which
requires that if a response matches the $i^{\textit{th}}$ method call
in the specification, then it must be preceded by at the first $i$
invocations of the specification.  This corresponds to
characterization of QQC given in \autoref{sec:qqc:2}.  In the case of
QQC, it is sufficient that a response by the $i^{\textit{th}}$ method
be preceded by any $i$ invocations, not necessarily the first $i$
invocations of the specification.

The proof of compositionality in \autoref{sec:lin:comp} is provided as
a warmup for the proof compositionality for QQC in
\autoref{sec:qqc:compose}.
\end{journal}

\subsection{First characterization: response to invocation}
\label{sec:lin:1}

Intuitively, linearizability requires that the response-to-invocation
order in an execution be respected by a specification trace.
To show that $\atr''$ is linearizable, it suffices to do the following
\begin{itemize}
\item Choose a specification trace $\btr$.
\item Choose an \emph{extension} $\atr'$ of $\atr''$ that closes the open
  calls in $\atr''$.   We say that $\atr'$ \emph{extends} $\atr''$ if
  \begin{enumerate*}
  \item if $\atr''$ is a prefix of $\atr'$, and
  \item all of the new events in $\atr'\setminus\atr''$ are ordered
    after all events of \ifconference{\emph}{opposite polarity} in
    $\atr''$
    (that is, calls after returns and returns after calls).
  \end{enumerate*}
  Let $\fextensions{\atr''}$ be the set of {extensions}\footnote{%
    \begin{math}
      \fextensions\ptr\eqdef \{\atr\;|\; \ptr\ltpre\atr
      \begin{array}[t]{>{\;}l<{\;}l}
        \land&
        \forall\zact!1\in\ptr.\; 
        \forall\zact?2\in\atr\setminus\ptr.\;\;
        \zact!1\lttrans{\ptr}\zact?2\\
        \land&
        \forall\zact?1\in\ptr.\; 
        \forall\zact!2\in\atr\setminus\ptr.\;\;
        \zact?1\lttrans{\ptr}\zact!2 \}
      \end{array}
    \end{math}}
  of $\atr''$.
\item Choose a renaming $\atr\ideq\atr'$ such that
  $\atr\permeq\btr$. \ifconference{Here $\ideq$ denotes
    equivalence up to renaming and $\permeq$ denotes equivalence up to
    permutation.}  This establishes that $\atr'$ is a
  permutation of $\btr$.  \ifjournal{Rather than carrying the permutation around
  in the definition, as usual in definitions of linearizability, we
  perform a renaming up front, once and for all.} The names are
  witness to the permutation.  \ifjournal{This works nicely, since our traces are
  indexed by names. Typically, linearizability is defined over
  strings, indexed by integers, so this technique is not available.}
\item Show that for every response $\zact!1$ and invocation
  $\zact?2$, if $\zact!1$ precedes $\zact?2$ in $\atr$
  (${\zact!1}\lttrans\atr{\zact?2}\,$), then the same must be true in
  $\btr$ (${\zact!1}\lttrans\btr{\zact?2}\,$).
\end{itemize}
\begin{journal}
Stated compactly, we have the following definition.

\begin{definition}
  \ifconference{\tinydisplayskip}
  Trace $\atr''$ \emph{linearizes} to
  $\btr$ if 
  $\exists\atr'\in\fextensions{\atr''}$.  
  $\exists\atr\ideq\atr'$.
  $\atr\permeq\btr$ and  
  \begin{displaymath}
    \forall\zact!1\!\in\atr.\;\;
    \forall\zact?2\!\in\atr.\;\;
    ({\zact!1}\lttrans\atr{\zact?2})
    \textimplies
    ({\zact!1}\lttrans\btr{\zact?2}).
  \end{displaymath}
  Trace set $\atrset$ \emph{linearizes} to $\btrset$ if
  $\forall\atr''\in\atrset$. $\exists\btr\in\btrset$. $\atr''$
  linearizes to $\btr$.
\end{definition}
\end{journal}

This definition differs from the traditional one in several small
details, \ifjournalelse{but is equivalent under reasonable assumptions.  The
differences are as follows.}{enumerated in the full paper. In particular, we allow $\atr'\in\fextensions{\atr''}$ to
  include {calls} that are not in $\atr''$, in addition to {returns}.}
\begin{journal}
\begin{itemize}
\item We do not require that specifications be sequential.  
\item We do not make requirements specific to threads.  A thread is
  simply a totally ordered sequence of actions, with the result that
  every pair of invocations must be separated by a response, and
  similarly for pairs of responses.  The fact that thread order is
  respected by linearizability follows from the general
  requirement that order from response to invocation must be
  respected.
\item In addition to \emph{returns}, we allow $\atr'\in\fextensions{\atr''}$ to
  include \emph{calls} that are not in $\atr''$.  Assuming that
  specifications are prefix-closed, this permissiveness is
  harmless. For every spec $\btr$ that includes the extra
  calls in a suffix, there is a corresponding spec
  $\btr'$ such that  $\btr\in\fextensions{\btr'}$ that does not
  include them (or their matching returns); if $\atr'$ with postpended
  call/return pairs linearizes to
  $\btr$, then $\atr'$ linearizes to $\btr'$\footnote{Informally, the
    argument is as follows: We must show that if $\atr'$ 
    is linearizable with the ability to add calls (and their matching
    returns) to the extension, then it is linearizable without that
    ability. Recall that any extension must be added \emph{after} the
    existing events.  For example, suppose
    \begin{math}
      \btr = 
      {\YINC1?0}  
      {\YINC2?1}
    \end{math}
    and
    \begin{math}
      \atr' =
      {\YINC1?0}.
    \end{math}
    Clearly $\atr'$ linearizes to $\btr$ if we postpend the missing call ${\YINC2?1}$.
    If we require that there be some 
    \begin{math}
      \btr' =
      {\YINC1?0},
    \end{math}
    then we can show $\atr'$ linearizes to $\btr'$.}.

\item We require that all incomplete calls remain in $\atr'$.
  Assuming that specifications are input-enabled, this restriction
  is harmless.  Input enabling simply means that an object cannot
  decide when it is called or with what parameters.  For every spec $\btr$ that does not include the
  extra calls, there is a corresponding spec
  $\btr'\in\fextensions{\btr}$ that does include them; 
  if $\atr'$ with the incomplete calls removed linearizes to $\btr$,
  then $\atr'$ linearizes to $\btr'$\footnote{Informally, the argument
    is as follows: We must show that if $\atr'$ 
    is linearizable with the ability to drop open calls, then it is
    linearizable without that ability.  For example, suppose
    \begin{math}
      \btr = 
      {\YINC1?0}  
      {\YINC2?1}
    \end{math}
    and
    \begin{math}
      \atr' =
      {\YINC1?0}  
      {\BINC3?2}
      {\YINC2?1}.
    \end{math}
    Clearly $\atr'$ linearizes to $\btr$ if we drop the open call ${\BINC3?2}$.
    If we require that there be some
    \begin{math}
      \btr' =
      {\YINC1?0}  
      {\YINC2?1}
      {\BINC3??}
      {\EINC3??},
    \end{math}
    (where \COLOR3{?} is any value) then we can show $\atr'$ linearizes to $\btr'$.}.
\end{itemize}
\end{journal}
We can refactor the definition slightly to pull it into the shape used
to define quiescent consistency and QQC.

\begin{definition}
  \label{def:lin:1}
  \ifconference{\tinydisplayskip}
  For traces $\atr$, $\btr$, we write $\atr\xsatlin\btr$ if $\atr\permeq\btr$ and
  for every prefix $\ptr\ltpre\atr$
  \begin{displaymath}
    \forall\zact!1\!\in\fevspol{}{\ptr}.\;\;
    \forall\zact?2\!\in\fevspol{}{\atr\setminus\ptr}.\;\;
    ({\zact!1}\lttrans\atr{\zact?2})
    \textimplies
    ({\zact!1}\lttrans\btr{\zact?2}).
  \end{displaymath}
  Then 
  \begin{math}
    (\atr''\satlin\btr)
    \eqdef
    (\exists\atr'\in\fextensions{\atr''}\!.\;\;
    \exists\atr\ideq\atr'\!.\;\;
    \atr\xsatlin\btr)\ifjournalelse{.}{,}
  \end{math}
  \begin{conference}
    \\and
    \begin{math}
      \mkern11mu
      (\atrset\satlin\btrset)
      \eqdef
      (\forall\atr''\in\atrset.\;\; 
      \exists\btr\in\btrset.\;\; 
      \atr''\satlin \btr).
    \end{math}
  \end{conference}
\end{definition}
\begin{journal}
\proofunskip
\begin{lemma}
  \label{thm:lin:prefix}
  $\atr$ linearizes to $\btr$ iff $\atr\satlin\btr$.
  \begin{compactproof}
    This is an immediate consequence of the definition of prefix.
  \end{compactproof}
\end{lemma}
\end{journal}

This characterization of linearizability requires that we look at
every way to \emph{cut} the trace $\atr$ into a prefix $\ptr$ and
suffix $\atr\setminus\ptr$.  We then look at the return
events in $\ptr$ and the call events in $\atr\setminus\ptr$ and ensure that the
order of events \emph{crossing the cut} is respected in $\btr$.  The
definitions are equivalent since we quantify over all possible cuts.

\ifjournalelse{As an example, c}{C}onsider the\ifjournal{ incrementing} counter specification from
\autoref{ex:intro}: 
\begin{math}
  {\YINC1?0}
  {\YINC3?1}
  {\YINC2?2}.
\end{math}
\ifjournal{For a completely concurrent trace, such as
\begin{math}
  {\BINC1??}
  {\BINC3??}
  {\BINC2??}
  {\EINC2!2}
  {\EINC3!1}
  {\EINC1!0}
\end{math}
linearizability is trivially satisfied since there is no cut that has
a return on the left and call on the right.}
The trace
\begin{math}
{\BINC3??}
{\BINC1??}
{\EINC3!1}
  {\BINC2??}
  {\EINC1!0}
  {\EINC2!2}
\end{math}
is \ifjournal{also} linearizable.  The interesting cut is 
\begin{math}
  {\BINC3??}
  {\BINC1??}
  {\EINC3!1}
\end{math}
which requires only that 
${\YINC3!1}$
precede
${\YINC2?2}$
in the specification.
By the same reasoning,
\begin{math}
{\BINC3??}
{\BINC2??}
\EINC3!1
  {\BINC1??}
  {\EINC2!2}
  {\EINC1!0},
\end{math}
is not linearizable, since it requires that 
${\YINC3!1}$
precede
${\YINC1?0}$.

\subsection{Second characterization: invocation to response}
\label{sec:lin:2}

Given a sequential specification, a trace is linearizable if every
return is preceded by the calls that come before it in specification
order.  This holds for \emph{operational} traces, in which all
events of opposite polarity are ordered.
\begin{conference}
  Operational traces correspond to those
  generated by a standard interleaving semantics.
\end{conference}
\begin{journal}
\begin{theorem}
  \label{thm:lin:2}
  Let $\btr$ be a sequential
  trace with name order
  \begin{math}
    (\xact?1\CC\xact!1\CC\xact?2\CC\xact!2\CC\ldots\CC\xact?n\CC\xact!n).
  \end{math}
  Let $\atr$ be an \operational{} trace such that $\atr\permeq\btr$.
  Then 
  \begin{displaymath}
    \atr\xsatlin\btr 
    \spq\textiff
    \forall j.\;
    \set{\xact?1\CC\ldots\CC \xact?j}
    \subseteq
    \setst{\xact?i}{
      {\xact?i}\lttrans\atr{\xact!j}
    }
  \end{displaymath}
  \begin{proof}
    Using the definition of linearizability and calculating, we have
    the following proof obligation.
    \begin{displaymath}
      (\forall i,j.\; {\xact!i}\lttrans\atr{\xact?j} \textimplies i<j)
      \spq\Leftrightarrow
      (\forall i,j.\; i\leq j \textimplies {\xact?i}\lttrans\atr{\xact!j})
    \end{displaymath}

    $(\Rightarrow)$ 
    Fix ${\xact!i}\lttrans\atr{\xact?j}$.  By way of contradiction,
    suppose $i\leq j$.  From the right implication we deduce that
    ${\xact?i}\lttrans\atr{\xact!j}$.  The resulting cycle,
    \begin{math}
      {\xact!i}
      \lttrans\atr
      {\xact?j}      
      \lttrans\atr
      {\xact!i}
    \end{math}
    contradicts the supposition that $\atr$ is a trace.  Therefore
    it must be that $i<j$ as required.
    
    $(\Leftarrow)$ 
    Fix $i\leq j$.  If $i=j$ the right implication holds by the
    definition of traces.  Suppose $i<j$.  By \operationality{}, either
    \begin{math}
      {\xact?i}\lttrans\atr{\xact!j}
    \end{math}
    or
    \begin{math}
      {\xact!j}\lttrans\atr{\xact?i}.
    \end{math}
    In the first case, the right implication holds.  In the second
    case, the left implication requires $j<i$, a contradiction.
  \end{proof}
\end{theorem}

Let us revisit the incrementing counter specification
\begin{math}
  {\YINC1?0}
  {\YINC3?1}
  {\YINC2?2}.
\end{math}
In the completely concurrent trace
\begin{math}
  {\BINC1??}
  {\BINC3??}
  {\BINC2??}
  {\EINC2!2}
  {\EINC3!1}
  {\EINC1!0}
\end{math}
all invocations precede all responses, and therefore linearizability is trivially satisfied.
The linearizability of
\begin{math}
  {\BINC3??}
  {\BINC1??}
  {\EINC3!1}
  {\BINC2??}
  {\EINC1!0}
  {\EINC2!2}
\end{math}
follows from the fact that 
${\EINC3!1}$ is preceded by both
${\BINC1??}$ and
${\BINC3??}$,
and the 
nonlinearizability of
\begin{math}
  {\BINC3??}
  {\BINC2??}
  {\EINC3!1}
  {\BINC1??}
  {\EINC2!2}
  {\EINC1!0},
\end{math}
follows from the fact that 
${\BINC1??}$ does not precede
${\EINC3!1}$.

\bigskip

The counting characterization also allows us to eliminate extension
from the top-level definition\footnote{This and similar corollaries
  for quiescent consistency and QQC are the only results in this paper
  that rely on the last of the four changes we have made in the
  definition on linearizability: ``We require that all incomplete
  calls remain in $\atr'$.''}.   
\end{journal}
\ifjournalelse{Recall that}{Define} $\atr\permlt\btr$ 
\ifjournalelse{indicates}{to mean} that $\atr$ is a subtrace of a permutation of $\btr$\ifjournalelse{.}{:
\begin{math}
  (\atr\permlt\btr) \eqdef
  (\exists \atr'.\;\; \atr\subseteq\atr' \permeq\btr).
\end{math}}

\ifjournalelse{\begin{corollary}}{\begin{theorem}}
  \label{thm:lin:2b}
  Let $\btr$ be a sequential
  trace with name order
  \begin{math}
    (\xact?1\CC\xact!1\CC\xact?2\CC\xact!2\CC\ldots\CC\xact?n\CC\xact!n).
  \end{math}
  Let $\atr\ifjournal{''}$ be an \operational{} trace such that $\atr\ifjournal{''}\permlt\btr$.
  Then 
  \begin{displaymath}
    \atr\ifjournal{''}\satlin\btr 
    \spq\textiff
    \forall \xact!j\in\atr\ifjournal{''}.\;
    \set{\xact?1\CC\ldots\CC \xact?j}
    \subseteq
    \setst{\xact?i}{
      {\xact?i}\lttrans{\atr\ifjournal{''}}{\xact!j}
    }
  \end{displaymath}
  \begin{proofsketch}
    $(\Rightarrow)$ Immediate from \autoref{thm:lin:2}.

    $(\Leftarrow)$ We need only show that there exists
    $\atr'\in\fextensions{\atr''}$ that satisfies the requirements.
    It suffices to take $\atr'= \atr'' ; (\btr\txminus\atr'')$.
  \end{proofsketch}
\ifjournalelse{\end{corollary}}{\end{theorem}}
\begin{journal}

This trick does not work for the primary definition, given in
\autoref{def:lin:1}.  For example, consider the counter trace
$\YINC1?5$.  This does not linearize to any counter specification, yet
it would be allowed if the requirement to extend $\atr''$ to a
permutation were dropped from \autoref{def:lin:1}.

\subsection{Compositionality}
\label{sec:lin:comp}

We re-prove one of the fundamental properties of linearizability:
compositionality \parencite{DBLP:journals/toplas/HerlihyW90}.  The
proof we give here is similar to the proof given for QQC in
\autoref{sec:qqc:compose}, in a simpler setting.

\begin{lemma}[\Operational{} traces]
  \label{thm:op}
  Suppose that $\atr$ is an \operational{} trace that imposes the
  following order.
  \begin{align*}
    \begin{tikzpicture}[xscale=1,yscale=-1]
      \node (b0) at (-.6,0) {$\xact?1$};
      \node (b3) at (1.6,0) {$\yact?1$};
      \node (b1) at (0,0) {$\xact?0$};
      \node (e1) at (0,1) {$\xact!0$};
      \node (b2) at (1,0) {$\yact?0$};
      \node (e2) at (1,1) {$\yact!0$};
      \draw [Query] (b1) to (e1);
      \draw [Query] (b2) to (e2);
      \draw [Query] (b1) to (e2);
      \draw [Query] (b2) to (e1);
      \draw [Query] (b0) to (e1);
      \draw [Query] (b3) to (e2);
    \end{tikzpicture}
  \end{align*}
  Then either ${\xact?1}\lttrans\atr{\yact!0}$ 
  or ${\yact?1}\lttrans\atr{\xact!0}$.
  \begin{compactproof}
    If neither holds, then, by \operationality{} we must have 
    both ${\yact!0}\lttrans\atr{\xact?1}$ 
    and ${\xact!0}\lttrans\atr{\yact?1}$, which results in the cycle
    ${\yact!0}\lttrans\atr{\xact?1}\lttrans\atr{\xact!0}\lttrans\atr{\yact?1}\lttrans\atr{\yact!0}$.
  \end{compactproof}
\end{lemma}

Recall from \autoref{sec:traces} that $(\interleave)$ denotes
interleaving and $(\txminus)$ denotes partial order difference.  To
split trace $\atr$ in ``half,'' it suffices to postulate the existence
of $\atr_1$ and $\atr_2$ such that $\atr_1=\atr\txminus\atr_2$ and
$\atr_2=\atr\txminus\atr_1$.
\begin{theorem}
  \label{thm:lin:comp}
  Let $\btr_1$ and $\btr_2$ be sequential traces.

  Let $\atr$, $\atr_1$ and $\atr_2$ be \operational{} traces such that
  $\atr_1=\atr\txminus\atr_2$ and
  $\atr_2=\atr\txminus\atr_1$. 

  For $i\in\set{1\CC2}$, suppose that each $\atr_i\xsatlin\btr_i$.
  
  Then there exists a sequential trace
  $\btr\in(\btr_1\interleave\btr_2)$ such that $\atr\xsatlin\btr$.
  \begin{proof}
    Without loss of generality, assume that $\feids{\btr_1}$ and
    $\feids{\btr_2}$ are disjoint.  Let the sequence of names in $\btr_1$
    be
    \begin{math}
      (\xact?1\CC\xact!1\CC\ldots\CC\xact?m\CC\xact!m)
    \end{math}
    and sequence of name in $\btr_2$ be
    \begin{math}
      (\yact?1\CC\yact!1\CC\ldots\CC\yact?n\CC\yact!n).
    \end{math}
    Applying \autoref{thm:lin:2} to the supposition
    $\atr_1\xsatlin\btr_1$, we have that $i\leq j$ implies
    ${\xact?i}\lttrans\atr{\xact!j}$, and similarly for the $\bact$s.

    Our aim is to construct a sequential interleaving of $\btr_1$ and
    $\btr_2$.  To do this, we construct a partial order over event pairs.
    Any interleaving consistent with the partial order will satisfy the
    conclusion of the theorem by construction.  For the elements of the
    partial order, let $\aact_i$ represent the pair $\xact?i\xact!i$ and
    let $\bact_k$ represent the pair $\yact?k\yact!k$.  Let the $\aact$s
    be totally ordered by subscript, corresponding to the fact that
    ${\xact?i}\lttrans\atr{\xact!j}$ whenever $i\leq j$, and similarly
    the $\bact$s.  Let there be a \emph{cross edge} from $\aact_i$ to
    $\bact_\ell$ if ${\xact!i}\lttrans\atr{\yact?\ell}$, and
    symmetrically from $\bact$s to $\aact$s.  Visually, we have an order
    such as the following.
    \begin{displaymath}
      \begin{tikzpicture}[xscale=1,yscale=-.7]
        \node (a1) at (0,0) {$\aact_1$};
        \node (a2) at (1,0) {$\aact_2$};
        \node (a3) at (2,0) {$\cdots$};
        \node (ai) at (3,0) {$\aact_i$};
        \node (a4) at (4,0) {$\cdots$};
        \node (aj) at (5,0) {$\aact_j$};
        \node (a5) at (6,0) {$\cdots$};
        \node (am) at (7,0) {$\aact_m$};
        \draw [query] (a1) to (a2);
        \draw [query] (a2) to (a3);
        \draw [query] (a3) to (ai);
        \draw [query] (ai) to (a4);
        \draw [query] (a4) to (aj);
        \draw [query] (aj) to (a5);
        \draw [query] (a5) to (am);
        \node (b1) at (0,1) {$\bact_1$};
        \node (b2) at (1,1) {$\bact_2$};
        \node (b3) at (2,1) {$\cdots$};
        \node (bk) at (3,1) {$\bact_k$};
        \node (b4) at (4,1) {$\cdots$};
        \node (bl) at (5,1) {$\bact_\ell$};
        \node (b5) at (6,1) {$\cdots$};
        \node (bn) at (7,1) {$\bact_n$};
        \draw [query] (b1) to (b2);
        \draw [query] (b2) to (b3);
        \draw [query] (b3) to (bk);
        \draw [query] (bk) to (b4);
        \draw [query] (b4) to (bl);
        \draw [query] (bl) to (b5);
        \draw [query] (b5) to (bn);
        \draw [query] (ai) to (bl);
        \draw [query] (b2) to (a1);
      \end{tikzpicture}
    \end{displaymath}
    The $\aact$-$\aact$ and $\bact$-$\bact$ edges go from $\pinp$ to $\pout$ in $\atr$,
    whereas the cross edges go from $\pout$ to $\pinp$.

    The proof obligation is to show that this order is acyclic, in which
    case it induces at least one interleaving.  We show that any cycle in
    the defined order corresponds to a cycle in $\atr$, contradicting the
    supposition that $\atr$ is a trace.  For there to be a cycle in the
    defined order, there must be $i<j$ and $k<\ell$, such that
    \begin{math} 
      {\xact?i} \lttrans\atr {\xact!j} \lttrans\atr
      {\yact?k} \lttrans\atr {\yact!\ell} \lttrans\atr {\xact?i}.
    \end{math}
    This contradicts the supposition that $\atr$ is a trace.
  \end{proof}
\end{theorem}
\end{journal}

\section{Quiescent Consistency}
\label{sec:qc}

Let $\ftropen{\atr}$ be the set of calls in $\atr$ that have
no matching return\footnote{%
\begin{math}
  \ftropen{\atr}\eqdef
  \setst{\albl\in\atr}{\fepol{\albl}=\pinp \;\land\; \mathord{\not\exists}\blbl\in\atr.\;\febrak{\blbl}=\feid{\albl}}
\end{math}}.
We say that trace $\atr$ is \emph{quiescent} if ${\ftropen{\atr}}=\emptyset$.
This notion of quiescence does not require that there be no
active thread, but only that there be no open calls.  
Thus, this notion of quiescence is compatible with libraries that
maintain their own thread pools.

The definition of quiescent consistency is similar to
\autoref{def:lin:1} of linearizability.  The difference lies in the
quantifier for the prefix $\ptr$: Whereas linearizability
quantifies over \emph{every} prefix, quiescent consistency only
quantifies over \emph{quiescent} prefixes.

\begin{definition}
  \label{def:qc}
  \ifconference{\tinydisplayskip}
  We write $\atr\xsatqc\btr$ if $\atr\permeq\btr$ and
  for any \emph{quiescent} prefix $\ptr\ltpre\atr$
  \begin{displaymath}
    \forall\zact!1\!\in\fevspol{}{\ptr}.\;\;
    \forall\zact?2\!\in\fevspol{}{\atr\setminus\ptr}.\;\;
    ({\zact!1}\lttrans\atr{\zact?2})
    \textimplies
    ({\zact!1}\lttrans\btr{\zact?2}).
  \end{displaymath}
  \begin{journal}
  Then 
  \begin{math}
    (\atr''\satqc\btr)
    \eqdef
    (\exists\atr'\in\fextensions{\atr''}\!.\;\;
    \exists\atr\ideq\atr'\!.\;\;
    \atr\xsatqc\btr).
  \end{math}
  \end{journal}
\end{definition}

\begin{conference}
  $(\satqc)$ is defined similarly to $(\satlin)$.
\end{conference}
Again let us revisit the counter specification from \autoref{ex:intro}:
\begin{math}
  {\YINC1?0}
  {\YINC3?1}
  {\YINC2?2}.
\end{math}
This notion of quiescent consistency places some constraints on the
system even when it has no nontrivial quiescent points.  For example,
the execution 
\begin{math}
  {\BINC1??}
  {\BINC3??}
  {\BINC2??}
  {\EINC2!3}
  {\EINC3!1}
  {\EINC1!0}
\end{math}
is not quiescently consistent with the given specification, since it
is not a permutation.  If one extends the execution to 
\begin{math}
  {\BINC1??}
  {\BINC3??}
  {\BINC2??}
  {\EINC2!3}
  {\EINC3!1}
  {\EINC1!0}
  {\YINC4?2}
\end{math}
and attempts to matches it against the specification
\begin{math}
  {\YINC1?0}
  {\YINC3?1}
  {\YINC4?2}
  {\YINC2?3},
\end{math}
quiescent consistency continues to fail: In the quiescent prefix
\begin{math}
  {\BINC1??}
  {\BINC3??}
  {\BINC2??}
  {\EINC2!3}
  {\EINC3!1}
  {\EINC1!0},
\end{math}
the order across the cut from ${\EINC2!3}$ to
${\BINC4?2}$ is not preserved in the specification.

For linearizability, \ifjournal{we argued that because specifications are
prefix-closed,} only responses need be included in the $\fextensions{}$
of a trace.  The same does not hold for quiescent consistency.  For
example, since
\begin{math}
  {\BINC2?2}
  {\YINC3?1}
  {\YINC1?0}  
  {\EINC2!2}
\end{math}
is quiescently consistent, its prefix
\begin{math}
  {\BINC2?2}
  {\YINC3?1}
\end{math}
should also be quiescently consistent.  However, there is no
specification trace that can be matched that does not include
\begin{math}
  {\YINC1?0}
\end{math}.
Therefore, it does not suffice merely to close the open call by adding
\begin{math}
  {\EINC2!2};
\end{math}
we must also include
\begin{math}
  {\BINC1?0}
\end{math}
and
\begin{math}
  {\EINC1!0}.
\end{math}

\begin{journal}
Compositionality (as expressed in \autoref{thm:lin:comp}) also holds
for quiescent consistency.  The proof is straightforward: any
quiescent point of $\atr_1\cup\atr_2$ is also a quiescent point for
each $\atr_i$; the two specifications may be interleaved arbitrarily
between these quiescent points.
\bigskip

\end{journal}
We now give a counting characterization of quiescent
consistency\ifjournal{ in the style of \autorefs{thm:lin:2} and
  \ref{thm:qqc:2}}.   
\begin{journal}
This characterization requires that if $\smash{\xact!j}$, the $j^{\text{th}}$ return in $\btr$, occurs in $\atr$, then there
must be at least $j$ calls contained in two sets: (1) the calls that
precede $\smash{\xact!j}$ in $\atr$, and (2) the calls that follow $\smash{\xact!j}$ in $\atr$ but
are ``quiescently concurrent'' --- that is, not separated by a
quiescent point.  
\end{journal}
\ifjournalelse{To capture the second set, we define}{Define}
\begin{math}
  \albl\lttransqc\atr\blbl
\end{math}
to mean that 
\begin{math}
  \albl\lttrans\atr\blbl
\end{math}
and there is no quiescent cut that
separates $\albl$ and $\blbl$.
\begin{journal}
\begin{definition}
  Define
  \begin{math}
    \albl\lttransqc\atr\blbl
  \end{math}
  to hold whenever 
  \begin{math}
    \albl\lttrans\atr\blbl
  \end{math}
  and there exists no quiescent prefix $\ptr\ltpre\atr$ such that 
  \begin{math}
    \albl\in\ptr
    \textand
    \blbl\in\atr\setminus\ptr.
  \end{math}
\end{definition}

\begin{theorem}
  \label{thm:qc:2}
  Let $\btr$ be a sequential
  trace with name order
  \begin{math}
    (\xact?1\CC\xact!1\CC\ldots\CC\xact?n\CC\xact!n).
  \end{math}
  Let $\atr$ be an \operational{} trace such that $\atr\permeq\btr$.
  Then 
  \begin{displaymath}
    \atr\xsatqc\btr
    \spq\textiff
    \forall j.\;
    \fcard{\set{\xact?1\CC\ldots\CC \xact?j}}    
    \leq
    \fcard{
      \setst{\xact?i}{{\xact?i}\lttrans\atr{\xact!j}}
      \cup
      \setst{\xact?i}{{\xact?i}\lttransqc\atr{\xact!j}}
    }
  \end{displaymath}
  \begin{proof}
    $(\Rightarrow)$ 
    Fix $j$ and
    let $\dtr$, $\ctr$ be the following disjoint sets.
    \begin{align*}
      \dtr &= \setst{\xact?i}{i\leq j \land \xact?i \lttrans\atr \xact!j}
      &
      \ctr&= \setst{\xact?i}{i\leq j \land \xact!j \lttransqc\atr \xact?i}
    \end{align*}
    If $i \leq j$ and $\xact!j \lttrans\atr \xact?i$, by
    \autoref{def:qc}, there is no quiescent cut that separates
    $\xact!j$ and $\xact?i$.  So, every $i \leq j$ that is not in
    $\dtr$ is in $\ctr$.  So, $|\dtr| + |\ctr| \geq j$.
    
    $(\Leftarrow)$ 
    Fix $\ptr$.    Fix $j=\max\setst{k}{\xact!k\in\ptr}$. 
    In order to show that the requirements of
    \autoref{def:qc} hold for every $\zact!1\in\ptr$, it suffices to
    show that they hold for $\xact!j$.

    We choose
    $\dtr$ and $\ctr$ as follows
    \begin{align*}
      \dtr &= \setst{\xact?i}{i\leq j \land \xact?i \lttrans\atr \xact!j}
      &
      \ctr &= \setst{\xact?i}{ \xact!j \lttransqc\atr \xact?i}
    \end{align*}
    Consider $\xact?i \in \ctr$.  Since $\ptr$ is a prefix of a
    quiescent cut, $\xact!i \in \ptr$.  By maximality of $j$, $i \leq
    j$.

    Since $|\dtr| + |\ctr| \geq j$, we deduce that $(\forall i \leq
    j.\; \xact?i \in \dtr \cup \ctr)$.  So, the requirements of
    \autoref{def:qc} hold for $\xact!j$.
  \end{proof}
\end{theorem}
\end{journal}

\ifjournalelse{\begin{corollary}}{\begin{theorem}}
  \label{thm:qc:2b}
  Let $\btr$ be a sequential
  trace with name order
  \begin{math}
    (\xact?1\CC\xact!1\CC\xact?2\CC\xact!2\CC\ldots\CC\xact?n\CC\xact!n).
  \end{math}
  Let $\atr\ifjournal{''}$ be an \operational{} trace such that $\atr\ifjournal{''}\permlt\btr$.
  Then 
  \begin{displaymath}
    \atr\ifjournal{''}\satqc\btr 
    \spq\textiff
    \forall \xact!j\in\atr\ifjournal{''}.\;
    \fcard{\set{\xact?1\CC\ldots\CC \xact?j}}
    \leq
    \fcard{
      \setst{\xact?i}{{\xact?i}\lttrans{\atr\ifjournal{''}}{\xact!j}}
      \cup
      \setst{\xact?i}{{\xact!j}\lttransqc\atr{\xact?i}}
    }
  \end{displaymath}
  \begin{proofsketch}
    Same as \autoref{thm:lin:2b}.
  \end{proofsketch}
\ifjournalelse{\end{corollary}}{\end{theorem}} 
\begin{conference}
If $\smash{\xact!j}$, the $j^{\text{th}}$ return in $\btr$, occurs in $\atr$, then there
must be at least $j$ calls contained in two sets: (1) the calls that
precede $\smash{\xact!j}$ in $\atr$, and (2) the calls that follow $\smash{\xact!j}$ in $\atr$ but
are ``quiescently concurrent'' --- that is, not separated by a
quiescent point.
\end{conference}

\begin{journal}
As noted in the introduction, if the sequence of interlocking calls
\begin{math}
  {\BINC1??}
  {\BINC2??}
  \smash{\EINC1!i} 
  {\BINC1??}
  \smash{\EINC2!j} 
  {\BINC2??}
  \smash{\EINC1!k}\allowbreak 
  {\BINC1??}
  \cdots,
\end{math}
never reaches quiescence, then the counter may return any natural
number for $\COLOR1i$, $\COLOR2j$ and $\COLOR1k$.  QQC reduces this
permissiveness by looking at every cut.  It remains less strict than
linearizability by loosening the requirement that \emph{every}
response-to-invocation across the cut be respected in the
specification.
\end{journal}

\section{Quantitative Quiescent Consistency}
\label{sec:qqc}

We provide three characterizations of QQC and prove their equivalence.
\begin{journalitemize}
\item \ifjournalelse{In \autoref{sec:qqc:1}, we define}{\autoref{def:qqc} defines}
  QQC in the style that we have defined linearizability and quiescent
  consistency, from response to invocation. 

\item \ifjournalelse{In \autoref{sec:qqc:2}, we give}{\autoref{thm:qqc:2} provides}
  a \emph{counting characterization} of QQC, which requires that if a
  response matches the $i^{\textit{th}}$ method call in the
  specification, then it must be preceded by at least $i$
  invocations. 
 
\item \ifjournalelse{In \autoref{sec:qqc:3}, we give}{\autoref{thm:qqc:3} provides}
  an operational characterization of QQC as a proxy between the
  concurrent world and an underlying sequential data structure.  
  \begin{journal}
    This can be seen a mix of flat combining
    \textcite{DBLP:conf/spaa/HendlerIST10} with speculation.
  \end{journal}
\end{journalitemize}
\begin{journal}
In \autoref{sec:qqc:compose}, we demonstrate that QQC is
compositional, in the sense of \textcite{DBLP:journals/toplas/HerlihyW90}.
Finally, in \autoref{sec:qqc:compare}, we compare QQC to the criterion
defined in \textcite{DBLP:conf/popl/HenzingerKPSS13}.
\bigskip
\end{journal}

To develop some intuition for the what is allowed by QQC, we give
some examples using the \texttt{$2$-Counter} from the introduction.
First we note that the capability given by an open call can be used
repeatedly, as in 
\begin{math}
  {\BINC2??}
  {\YINC1?1}
  {\YINC3?0}
  {\YINC1?3}
  {\YINC3?2}
  {\YINC1?5}
  \YINC3?4
  {\EINC2!6}.
\end{math}
The open call ${\BINC2??}$ enables the inversion of
${\YINC3?0}$ with ${\YINC1?1}$ and also of
${\YINC3?2}$ with ${\YINC1?3}$.

Alternatively, multiple open calls may be accumulated to create 
an trace with events that are arbitrarily far off, as in 
\begin{math}
  {\BINC2??}
  {\YINC1?1}
  {\BINC2??}
  {\YINC1?3}
  {\BINC2??}
  {\YINC1?5}
  {\BINC2??}
  {\YINC1?7}
  {\YINC1?0}
  {\EINC2!2}
  {\EINC2!4}
  {\EINC2!6}
  {\EINC2!8}.
\end{math}
Note that ${\YINC1?0}$ \emph{follows} ${\YINC1?7}$ in this
execution!  It is worth emphasizing that the order between these
actions is observable to the outside: a single thread can call
\texttt{get\-And\-Inc\-re\-ment} and get $7$, then subsequently call
\texttt{get\-And\-Inc\-re\-ment} and get $0$.  Such behaviors are a
hallmark of nonlinearizable data structures.
In general, an \texttt{$N$-\allowbreak{}Counter} can give results that
are $k\times{}N$ off of the expected value, where $k$ is the maximum
number of open calls and $N$ is the width of the counter.  There is no
way to bound the behavior of this counter, as
in \parencite{DBLP:conf/popl/HenzingerKPSS13}, without also bounding
the amount of concurrency, as
in \parencite{DBLP:conf/opodis/AfekKY10}.

It is also possible for open calls to overlap in nontrivial ways.
The trace
\begin{math}
  \BINC2??
  \YINC1?1
  \BINC3??
  \YINC1?0
  \EINC2!3
  \YINC2?2
  \EINC3!4
\end{math}
is QQC.
Here, the first
$\BINC2??$
justifies the out-of-order execution of
$\YINC1?1$ and
$\YINC1?0$.
The subsequent
$\BINC3??$
justifies an inversion of the previous justifier, 
namely
$\YINC2!3$
and
$\YINC2?2$.
A similar example is
\begin{math}
  \BINC3??
  \YINC2?1
  \BINC2??
  \YINC1?0
  \EINC2!3
  \YINC1?2
  \EINC3!4.
\end{math}

Finally, we note that the stack execution
\begin{math}
  \BPUT3?c
  \YGET1.a
  \YPUT2.a
  \EPUT3!c
\end{math}
is QQC with respect to the specification
\begin{math}
  \YPUT2.a
  \YGET1.a
  \BPUT3?c
  \EPUT3!c\!.
\end{math}
This follows from exactly the kind of reasoning that we have done for
the counter.  For the counter this simply means that we are seeing an
integer value early, but for a stack holding pointers, it means that
we can potentially see a pointer before it has been allocated!  To prevent such executions, causality can be
specified as a relation from calls to returns, 
consistent with specification order: \ifjournalelse{An implementation}{A} trace is
\emph{causal} if it respects the specified causality relation.  We
have elided causality from the definition of QQC because it is
orthogonal and can be enforced independently.

\subsection{First characterization: response to invocation}
\label{sec:qqc:1}

Linearizability requires that for \emph{every} cut, \emph{all}
response-to-invocation order crossing the cut must be respected in the
specification.  Quiescent consistency limits attention to
\emph{quiescent} cuts.  QQC restores the quantification over {every}
cut, but relaxes the requirement to match all response-to-invocation
order crossing the cut.  When checking response-to-invocation pairs
across the cut, QQC allows some invocations to be ignored.  How many?

One constraint comes from our desire to refine quiescent consistency.
For quiescent cuts, we cannot drop any invocations, since quiescent
consistency does not.  As a first attempt at a definition, we may take
the number of dropped invocations at any cut to be bounded by
$\fcard{\ftropen{\ptr}}$.  This criterion would allow both of the traces
\begin{journal}
\begin{align*}
  {\BINC2?2}
  {\YINC3?1}
  {\YINC1?0}  
  {\EINC2!2}
  &&
  {\BINC1?0}
  {\YINC2?2}  
  {\YINC3?1}
  {\EINC1!0}
\end{align*}
\end{journal}
\begin{conference}
\begin{math}
  {\BINC2?2}
  {\YINC3?1}
  {\YINC1?0}  
  {\EINC2!2}
\end{math}
and
\begin{math}
  {\BINC1?0}
  {\YINC2?2}  
  {\YINC3?1}
  {\EINC1!0}
\end{math}
\end{conference}
in \autoref{ex:intro}.
In each case, the interesting cut splits the trace in half, with one
open call and one completed.   
In the first trace, we can ignore
\begin{math}
  {\BINC1?0}
\end{math}
in the suffix, and in the second trace, we can ignore
\begin{math}
  {\BINC3?1}
\end{math}
in the suffix; thus, both\ifjournal{ traces} are allowed\ifjournal{ by this first attempt}.
However, in the second trace, the first call completed is two steps in the
future, even though there is only one concurrent action.  In the first
trace this does not happen.
The difference can be seen by looking not only at the number of open
calls, but also at \emph{which} calls are open.  In the first trace
we have
\begin{math}
  {\BINC2?2}
\end{math}
before
\begin{math}
  {\EINC3!1}\!,
\end{math}
and in the second, we have
\begin{math}
  {\BINC1?0}
\end{math}
before
\begin{math}
  {\EINC2!2}\!.
\end{math}
We say that
\begin{math}
  {\BINC2?2}
\end{math}
is \emph{early} for 
\begin{math}
  {\EINC3!1}\!,
\end{math}
since it does {not} precede
\begin{math}
  {\EINC3!1}
\end{math}
in the specification,
whereas
\begin{math}
  {\BINC1?0}
\end{math}
is not early for
\begin{math}
  {\EINC2!2}\!,
\end{math}
since it \emph{does} precede
\begin{math}
  {\EINC2!2}\!.
\end{math}
We restrict our attention to calls that are both open and early
with respect to the response of interest.

Given a specification $\btr$ and a response $\zact!1\in\btr$, none of the actions
in the $\btr$-down\-closure of $\zact!1$ could possibly be early for
$\zact!1$; any other action could be.  Thus, the actions in
\begin{math}
  {\ftropen{\ptr}}\setminus\allowbreak(\fedown\btr{\zact!1})
\end{math}
are both open and early for $\zact!1$.  This leads us to the following
definition.  (In \ifjournalelse{\autoref{sec:qqc:2}}{the full
  paper}, we show that for sequential specifications,
we can swap the quantifiers $(\exists\ctr)$ and $(\forall\zact!1)$,
pulling out the existential.)
\begin{definition}
  \label{def:qqc}
  \ifconference{\tinydisplayskip}
  We write $\atr\xsatqqc\btr$ if $\atr\permeq\btr$ and
  for any prefix $\ptr\ltpre\atr$
  \begin{displaymatharrayl}
    \forall\zact!1\!\in\fevspol{}{\ptr}.\;\;
    \exists\ctr\subseteq\atr.\;\fcard\ctr \leq \fcard{\ftropen{\ptr}\setminus(\fedown\btr{\zact!1})}.\;\;\\\QQUAD
    \forall\zact?2\!\in\fevspol{}{((\atr\setminus\ptr)\setminus\ctr)}.\;\;
    ({\zact!1}\lttrans\atr{\zact?2})
    \textimplies
    ({\zact!1}\lttrans\btr{\zact?2}).
  \end{displaymatharrayl}
  \begin{journal}
  Then 
  \begin{math}
    (\atr''\satqqc\btr)
    \eqdef
    (\exists\atr'\in\fextensions{\atr''}\!.\;\;
    \exists\atr\ideq\atr'\!.\;\;
    \atr\xsatqqc\btr).
  \end{math}
  \end{journal}
\end{definition}
\begin{journal}

In this definition, it is safe to restrict attention to sets $\ctr$
consisting only of input events that are concurrent with the open
calls.  We do not impose these restrictions explicitly because they
are not necessary.  Choosing outputs does not add any flexibility,
effectively wasting an open call.  Non-concurrent calls will be
revealed by the prefix in which the call is closed.
\end{journal}
\ifconference{\vspace{-.5\baselineskip}}
\begin{conference}
  As before, $(\satqqc)$ is defined by analogy to $(\satlin)$.
\end{conference}
\begin{theorem}
  \label{thm:lin:qqc:qc}
  \begin{math}
    (\satlin) \subset (\satqqc) \subset (\satqc)
  \end{math}
  \begin{proof}
    Containment is immediate from the definitions, always taking
    $\ctr=\trempty$ for QQC.  To see that the containment is proper,
    consider the incrementing counter specification from 
    \autoref{ex:intro}, 
    \begin{math}
      {\YINC1?0}
      {\YINC2?1}
      {\YINC3?2}.
    \end{math}
    With respect to this specification,
    \begin{math}
      {\BINC3??}
      {\BINC2??}
      {\EINC2!1}
      {\BINC1??}
      {\EINC1!0}
      {\EINC3!2}
    \end{math}
    is QQC but not linearizable
    \begin{math}
      {\BINC1??}
      {\BINC3??}
      {\EINC3!2}
      {\BINC2??}
      {\EINC2!1}
      {\EINC1!0}
    \end{math}
    is quiescently consistent but not QQC.
  \end{proof}
\end{theorem}

\begin{journal}
  If there are no overlapping calls, then 
  $(\satlin)$, $(\satqqc)$ and $(\satqc)$ coincide.

  Recall the definition of sequential consistency: Two traces are
  sequentially consistent if they are equal on every projection to a
  single thread.  Two define this in our formalism, we require that
  labels include a thread identifier, and that projecting a trace to a
  single thread gives a is totally ordered subtrace.  For operational
  traces, linearizability refines sequential consistency.
  
  Like quiescent consistency, QQC is incomparable to sequential
  consistency: In \autoref{ex:intro}, the second and third traces are
  QQC but not sequentially consistent.  In the other direction,
  History $H_7$ of \parencite[\textsection
  3.3]{DBLP:journals/toplas/HerlihyW90}, is sequentially consistent
  but not QQC.  The same example is given in Fig 3.8
  of \parencite{HS08}.  The argument
  in \parencite{DBLP:journals/toplas/HerlihyW90,HS08} concerns
  linearizability, but the example has no overlapping calls and
  therefore applies equally to QQC.

\subsection{Second characterization: counting invocations}
\label{sec:qqc:2}
\end{journal}

Given the subtlety of \autoref{def:qqc}, it may be surprising that QQC
has the following simple characterization for sequential
specifications.  
\begin{journal}
\begin{theorem}
  \label{thm:qqc:2}
  Let $\btr$ be a sequential
  trace with name order
  \begin{math}
    (\xact?1\CC\xact!1\CC\ldots\CC\xact?n\CC\xact!n).
  \end{math}
  Let $\atr$ be an \operational{} trace such that $\atr\permeq\btr$.
  Then 
  \begin{displaymath}
    \atr\xsatqqc\btr
    \spq\textiff
    \forall j.\;
    \fcard{\set{\xact?1\CC\ldots\CC \xact?j}}
    \leq
    \fcard{\setst{\xact?i}{
        {\xact?i}\lttrans\atr{\xact!j}
      }}
  \end{displaymath}
  \begin{compactproof}
    $(\Rightarrow)$ 
    Fix $j$, let $\ptr=\fedown\atr{\xact!j}$, and
    let $\dtr$, $\ctr'$, $\etr$ be the following disjoint sets.
    \begin{align*}
      \dtr &= \setst{\xact?i}{i\leq j \land \xact?i \lttrans\atr \xact!j}
      \\
      \ctr' &= \setst{\xact?i}{i\leq j \land \xact?i \not\lttrans\atr \xact!j}
      = \setst{\xact?i}{i\leq j \land \xact!j \lttrans\atr \xact?i}
      &&\text{(by \operationality{})}
      \\
      \etr &= \setst{\xact?i}{i> j \land \xact?i \lttrans\atr \xact!j}
      \supseteq \ftropen{\ptr}\setminus(\fedown\btr{\xact!j})
      &&\text{(by calculation)}
    \end{align*}
    Note that 
    \begin{math}
      \dtr\cup\etr=
      \setst{\xact?i}{{\xact?i}\lttrans\atr{\xact!j}};
    \end{math}
    therefore it suffices to show that 
    \begin{math}
      \fcard{\dtr\cup\etr}\geq j.
    \end{math}

    For every event in $\xact?i\in\ctr'$ we have that $i\leq j$ and
    therefore ${\xact!j}\lttrans\atr{\xact?i}$ and
    ${\xact!j}\not\lttrans\btr{\xact?i}$. Hence the set $\ctr$ chosen
    in \autoref{def:qqc} must include $\ctr'$.  From
    \autoref{def:qqc}, we have that
    \begin{math}
      \fcard{\ctr}
      \leq
      \fcard{\ftropen{\ptr}\setminus(\fedown\btr{\xact!j})}.
    \end{math}
    Since
    $\ctr'\subseteq\ctr$ and
    $\ftropen{\ptr}\setminus(\fedown\btr{\xact!j}) \subseteq \etr$, 
    we have 
    \begin{math}
      \fcard{\ctr'}
      \leq
      \fcard{\etr}.
    \end{math}
    Since
    \begin{math}
      \fcard{\dtr \cup \ctr'}=j,
    \end{math}
    we have
    \begin{math}
      \fcard{\dtr \cup \etr}\geq j,
    \end{math}
    as required.

    $(\Leftarrow)$ 
    Fix $\ptr$.  
    Following the argument given in the proof of \autoref{thm:qqc:exists},
    in order to show that the requirements of
    \autoref{def:qqc} hold for every $\zact!1\in\ptr$, it suffices to
    show that they hold for $\xact!j$, where let
    $j=\max\setst{k}{\xact!k\in\ptr}$.

    Fix $j=\max\setst{k}{\xact!k\in\ptr}$. We now show that the
    requirements of \autoref{def:qqc} hold for $\xact!j$.  We choose
    $\dtr$, $\ctr$ and $\etr$ as before.
    \begin{align*}
      \dtr &= \setst{\xact?i}{i\leq j \land \xact?i \lttrans\atr \xact!j}
      \\
      \ctr &= \setst{\xact?i}{i\leq j \land \xact?i \not\lttrans\atr \xact!j}
      = \setst{\xact?i}{i\leq j \land \xact!j \lttrans\atr \xact?i}
      \\
      \etr &= \setst{\xact?i}{i> j \land \xact?i \lttrans\atr \xact!j}
      \subseteq \ftropen{\ptr}\setminus(\fedown\btr{\xact!j})
    \end{align*}
    To see that $\etr \subseteq \ftropen{\ptr}$, consider that if
    \begin{math}
      \xact?i\in\etr
    \end{math}
    then
    \begin{math}
      \xact!i\not\in\ptr;
    \end{math}
    otherwise $j\not=\max\setst{k}{\xact!k\in\ptr}$.  By the second
    characterization of $\ctr$ above (which follows from \operationality{}),
    \begin{math}
      \forall\xact?i\!\not\in\fevspol{}{\ctr}.\;\;
      ({\xact!j}\lttrans\atr{\xact?i}) \textimplies
      j<i.
    \end{math}
    Thus, to establish the result it suffices to show that 
    \begin{math}
      \fcard{\ctr}\leq \fcard{\ftropen{\ptr}\setminus(\fedown\btr{\xact!j})}.
    \end{math}
    By assumption, 
    \begin{math}
      \fcard{\dtr \cup \etr}\geq j.
    \end{math}
    Since
    \begin{math}
      \fcard{\dtr \cup \ctr}=j,
    \end{math}
    we have
    \begin{math}
      \fcard{\ctr}\leq \fcard{\etr}
    \end{math}
    and therefore
    \begin{math}
      \fcard{\ctr}\leq \fcard{\ftropen{\ptr}\setminus(\fedown\btr{\xact!j})}
    \end{math}
    as required.
  \end{compactproof}
\end{theorem}
\end{journal}
\ifjournalelse{\begin{corollary}}{\begin{theorem}}
  \ifjournalelse{\label{thm:qqc:2b}}{\label{thm:qqc:2}}
  Let $\btr$ be a sequential
  trace with name order
  \begin{math}
    (\xact?1\CC\xact!1\CC\xact?2\CC\xact!2\CC\ldots\CC\xact?n\CC\xact!n).
  \end{math}
  Let $\atr\ifjournal{''}$ be an \operational{} trace such that $\atr\ifjournal{''}\permlt\btr$.
  Then 
  \begin{displaymath}
    \atr\ifjournal{''}\satqqc\btr 
    \spq\textiff
    \forall \xact!j\in\atr\ifjournal{''}.\;
    \fcard{\set{\xact?1\CC\ldots\CC \xact?j}}
    \leq
    \fcard{\setst{\xact?i}{
      {\xact?i}\lttrans{\atr\ifjournal{''}}{\xact!j}
    }}
  \end{displaymath}
  \begin{proofsketch}
    Same as \autoref{thm:lin:2b}.
  \end{proofsketch}
\ifjournalelse{\end{corollary}}{\end{theorem}}
\ifjournal{\par}
This characterization provides a simple method for calculating whether
a trace is QQC.  For example, the trace
\begin{math}
  \BINC3??
  \YINC2?1
  \BINC2??
  \YINC1?0
  \EINC2!3
  \YINC1?2
  \EINC3!4
\end{math}
is QQC since
$\EINC2!1$ is preceded by two calls,
$\EINC1!0$, $\EINC2!3$ by four, and
$\EINC1!2$, $\EINC3!4$ by five.
The trace
\begin{math}
  {\BINC3??
  \YINC2?1
  \BINC2??
  \EINC2!3
  \YINC1?0
  \YINC1?2
  \EINC3!4}
\end{math}
is not QQC since $\EINC2!3$ is only preceded by three calls, yet it is
the fourth call in the specification.
\begin{journal}

For sequential specifications, we can also simplify \autoref{def:qqc}
by exchanging the quantifiers $(\exists\ctr)$ and $(\forall\zact!1)$,
pulling out the existential.  
\begin{lemma}
  \label{thm:qqc:exists}
  Let $\btr$ be a sequential
  trace with name order
  \begin{math}
    (\xact?1\CC\xact!1\CC\ldots\CC\xact?n\CC\xact!n).
  \end{math}
  Let $\atr$ be an \operational{} trace such that $\atr\permeq\btr$.
  Fix $\ptr\ltpre\atr$.
  Then the displayed requirement of \autoref{def:qqc} is equivalent to
  \begin{displaymatharrayl}
    \exists\ctr\subseteq\atr.\,\fcard\ctr \leq 
    \fcard{\fearlyopen{\btr}{\ptr}}.\;\;\\\QQUAD\;
    \forall\zact!1\!\in\fevspol{}{\ptr}.\;\;
    \forall\zact?2\!\in\fevspol{}{((\atr\setminus\ptr)\setminus\ctr)}.\;\;
    ({\zact!1}\lttrans\atr{\zact?2})
    \textimplies
    ({\zact!1}\lttrans\btr{\zact?2}),
  \end{displaymatharrayl}  
  where
  \begin{math}
    \fearlyopen{\btr}{\ptr} \eqdef
    \setst{\zact?2\in\ftropen{\ptr}}{\not\exists\zact!1\in\ptr.\; \zact?2\lttrans\btr\zact!1}.
  \end{math}
  \begin{proof}
    (\ref{thm:qqc:exists} $\Rightarrow$ \ref{def:qqc}) Immediate.

    (\ref{def:qqc} $\Rightarrow$ \ref{thm:qqc:exists}) Consider the
    proof of the reverse direction ($\Leftarrow$) in the
    \autoref{thm:qqc:2}.  An examination of the proof shows that the
    open calls constructed satisfy the more stringent requirements of
    \ref{thm:qqc:exists}.  In fact, the proof of \ref{thm:qqc:2} shows
    that (\ref{thm:qqc:2} $\Rightarrow$ \ref{thm:qqc:exists}).  The
    result follows since the forward direction of \ref{thm:qqc:2}
    shows that (\ref{def:qqc} $\Rightarrow$ \ref{thm:qqc:2}).
  \end{proof}
\end{lemma}
For full concurrent specifications and implementations, we suspect
that \autoref{thm:qqc:exists} fails.  
(To get a sense of the issues, consider a specification that orders
$\COLOR1a\xquery\COLOR1c$ and 
$\COLOR2b\xquery\COLOR2d$, and an implementation that executes
$\COLOR1a\xquery\COLOR2d$ and 
$\COLOR2b\xquery\COLOR1c$.)
In this paper, however, all of
our results concern sequential specifications and operational implementations.
\end{journal}

\subsection{Third characterization: speculative flat combining}
\label{sec:qqc:3}

Our third characterization of QQC describes how QQC affects an
arbitrary sequential data structure, using a \emph{proxy} that
generates QQC traces from an underlying sequential implementation.
\begin{journal}
The proxy is \emph{sound}, in that every trace that it accepts is QQC,
and \emph{complete}, in that it generates every operational trace that
is QQC with respect to the sequential data structure.
\par
\end{journal}
This characterization of QQC incorporates \emph{speculation} into flat
combining \parencite{DBLP:conf/spaa/HendlerIST10}.
\ifjournalelse{\emph{Flat combining} is a technique for implementing
  concurrent data structures using sequential ones by introducing a
  mediator between the concurrent world and the sequential data
  structure.  As for speculation, w}{W}e push the obligation to
predict the future into the underlying sequential object, with must
conform to the following interface.
\begin{lstlisting}
interface Object {
   method run(i:Invocation):Response;
   method predict():Invocation;  }
\end{lstlisting}
The \texttt{run} method passes invocations to the underlying
sequential structure and returns the appropriate response.  
The \texttt{predict} method is an oracle that guesses the
invocations that are to come in the future.  It is the use of
\texttt{predict} that makes our code speculative.

\begin{journal}
Given an \texttt{Object} \texttt{o}, the proxy is defined as follows.
\end{journal}
\begin{figure}
\begin{lstlisting}
class #\ctype{QQCProxy<o:Object>}# {
   field called:#\ctype{ThreadSafeMultiMap<Invocation,Semaphore>}# = [];
   field returned:#\ctype{ThreadSafeMap\ \ \ <Semaphore, Response>}#  = [];
   method run(i:Invocation):Response { // proxy for external access to o
     val m:Semaphore = [];
     called.add(i, m);
     m.wait();
     return returned.remove(m); }
   thread { // single thread to interact with o
     val received:#\ctype{MultiMap<Invocation,Semaphore>}# = [];
     val executed:#\ctype{MultiMap<Invocation,Response>}#  = [];
     repeatedly choose {
        choice if called.notEmpty() {
           received.add(called.removeAny());
           val i:Invocation = o.predict();
           val r:Response   = o.run(i);
           executed.add(i, r); }
        choice if exists i in received.keys() intersect executed.keys() {
           val m:Semaphore = received.remove(i);
           val r:Response  = executed.remove(i);
           returned.add(m, r);
           m.signal(); } } } }
\end{lstlisting}
\caption{QQC Proxy}
\label{fig:proxy}
\end{figure}

The code for the proxy is given in \autoref{fig:proxy}. Communication between the implementation threads and the underlying
\texttt{Object} is mediated by two maps.
When a thread would like to interact with the \texttt{Object}, it
creates a semaphore, registers \ifjournalelse{the semaphore}{it} in
\texttt{called} and waits\ifjournal{ on the semaphore}.  Upon awakening,
the thread removes the result from \texttt{returned} and returns.

The \texttt{Object} is serviced by a single \emph{proxy} thread which
loops forever making one of two nondeterministic choices. The proxy
keeps two private maps.  Upon receiving an invocation in \texttt{called}, the
proxy moves the invocation from \texttt{called} to \texttt{received}.
Rather than executing the received invocation, the proxy asks the
oracle to predict an arbitrary invocation \texttt{i} and executes that
instead, placing the result in \texttt{executed}.  Once a invocation
is both \texttt{received} and \texttt{executed}, it may become
\texttt{returned}.

At the beginning of this section, we noted that the stack execution 
\begin{math}
  {\BPUT3?c
  \YGET1.a
  \YPUT2.a
  \EPUT3!c}
\end{math}
is QQC with respect to the specification
\begin{math}
  \YPUT2.a
  \YGET1.a
  \BPUT3?c
  \EPUT3!c\!.
\end{math}
How can such a trace possibly be generated?  The execution of the
proxy proceeds as follows.
Upon receipt of ${\BPUT3?c}$\!, the proxy executes
$\BPUT2?a$\!, storing response $\EPUT2!a$\!\!.
Upon receipt of ${\BGET1?a}$\!\!, the proxy executes
${\BGET1?a}$\!\!, storing response $\EGET1!a$.
At this point $\BGET1?a\EGET1!a$ can return.
Upon receipt of $\BPUT2?a$\!, the proxy executes 
$\BPUT3?c$\!, storing response $\EPUT3!c$\!\!.
At this point both $\BPUT2?a\EPUT2!a$ 
and $\BPUT3?c\EPUT3!c$ can return.

Such noncausal behaviors can be eliminated by requiring when a pop is
executed, a corresponding push must have been received.  The prior
execution is invalidated since $\YPUT2.a$ is not received when
$\YGET1.a$\! returns.  However, nonlinearizable behaviors are still
allowed.  For example
\begin{math}
  \BPUT3?c
  \YPUT1.a
  \YPUT2.b
  \EPUT3!c
  \YGET1.a
  \YGET2.b
\end{math}
is generating by predicting $\YPUT2.b$\ifjournal{ before $\YPUT1.a$}\!\!.

\begin{theorem}
  \label{thm:qqc:3}
  The concurrent proxy is sound for QQC with respect to the underlying
  \texttt{Object}.  It is also complete for \operational{} traces.
  \begin{proof}
    For soundness, note that proxy maintains the invariant that the
    sizes of \texttt{received} and \texttt{executed} are equal,
    and therefore the number of returned calls can never exceed the
    number that has been received.  In addition, the number of things
    added to \texttt{received} always exceeds the number added to
    \texttt{returned}.

    For completeness, suppose that trace $\atr\xsatqqc\btr$ and let
    the sequence of names in $\btr$ be
    \begin{math}
      (\xact?1\CC\xact!1\CC\ldots\CC\xact?m\CC\xact!m).
    \end{math}
    Consider any total order on the events of $\atr$ that is consistent with the
    order already present in $\btr$.  Let 
    \begin{math}
      (\yact?1\CC\ldots\CC\yact?m)
    \end{math}
    be the order on the call actions in this total order.  When
    $\yact?i$ arrives, add $\yact?i$ to \texttt{received} and execute
    $\xact?i$, placing $\xact!i$ into \texttt{executed}.  From
    \autoref{thm:qqc:2} we know that whenever a response is
    required, there will be enough prior invocations so that the
    required response will be found in \texttt{executed}.
  \end{proof}
\end{theorem}
\begin{conference}
\vspace{-.5\baselineskip}
\par In the full paper, we show that the elimination-tree stack of
\parencite{DBLP:journals/mst/ShavitT97} and increment-only counter
of \parencite{DBLP:journals/jacm/AspnesHS94} are QQC.  The 
characterizations of QQC also allow us to predict the QQC behavior of
other data structures, such as a queues, even if no implementation is known.
The following examples, from \textcite{Ali}, allow a useful comparison
with \parencite{DBLP:conf/popl/HenzingerKPSS13}.  

To see that QQC makes distinctions  not
found in \parencite{DBLP:conf/popl/HenzingerKPSS13}, consider the
two stack traces
\begin{math}
  \BPUT3?c
  \YPUT1.a
  \YPUT2.b
  \YGET4.a
  \EPUT3!c
\end{math}
and
\begin{math}
  \BPUT3?c
  \YPUT1.a
  \YPUT2.b
  \EPUT3!c
  \YGET4.a\!.
\end{math}
In the framework of \parencite{DBLP:conf/popl/HenzingerKPSS13}, these
are both $1$ out-of-order (when $\COLOR1a$ is popped, at least
$\COLOR2b$ must be above $\COLOR1a$ on the stack).  However, only the
first is QQC.

In the other direction, the queue execution
\begin{math}
  \BPUT3?a
  \YPUT1.{b_1}
  \YPUT1.{b_1}
  \cdots
  \YPUT1.{b_n}
  \YPUT2.c
  \YGET4.c
  \EPUT3!a
\end{math}
is QQC with respect to the queue specification
\begin{math}
  \YPUT2.c
  \YPUT1.{b_1}
  \YPUT1.{b_1}
  \cdots
  \YPUT1.{b_n}
  \YGET4.c
  \BPUT3?a
  \EPUT3!a\!\!.
\end{math}
In the framework of \parencite{DBLP:conf/popl/HenzingerKPSS13}, this would be $n$ out-of-order because at least all
$\COLOR1{b_i}$'s should be in the queue before $\COLOR2c$ is inserted
into the queue; the removal of $\COLOR2c$ from the queue must happen when there are $n$ elements
ahead of $\COLOR2c$ in the queue.

\medskip
Finally, we prove compositionality for QQC. Let $\txminus$ denote partial order difference.
\end{conference}

\begin{journal}
\subsection{Compositionality}
\label{sec:qqc:compose}

We now prove compositionality for QQC, following the proof for
linearizability in \autoref{thm:lin:comp}.  Below, we give some
examples of the construction given in the proof, which is more complex
than the one required for linearizability.  Recall that
($\txminus$) denotes partial order difference.
\end{journal}
\begin{theorem}
  \label{thm:qqc:comp}
  Let $\btr_1$ and $\btr_2$ be sequential traces.

  Let $\atr$, $\atr_1$ and $\atr_2$ be \operational{} traces such that
  $\atr_1=\atr\txminus\atr_2$ and
  $\atr_2=\atr\txminus\atr_1$. 

  For $i\in\set{1\CC2}$, suppose that each $\atr_i\xsatqqc\btr_i$.
  
  Then there exists a sequential trace
  $\btr\in(\btr_1\interleave\btr_2)$ such that $\atr\xsatqqc\btr$.
  \begin{proofkeep}
    \ifjournalelse{As in the proof of \autoref{thm:lin:comp}, assume $\feids{\btr_1}$
    and $\feids{\btr_2}$ are disjoint, and l}{Assume that the names in ${\btr_1}$
    and ${\btr_2}$ are disjoint. L}et the sequence of names
    in $\btr_1$ be
    \begin{math}
      (\xact?1\CC\xact!1\CC\ldots\CC\xact?m\CC\xact!m)
    \end{math}
    and sequence of names in $\btr_2$ be
    \begin{math}
      (\yact?1\CC\yact!1\CC\ldots\CC\yact?n\CC\yact!n).
    \end{math}
    Applying \autoref{thm:qqc:2} to the supposition
    $\atr_1\xsatlin\btr_1$, we have that 
    \begin{math}
      j\leq\fcard{\setst{\xact?i}{{\xact?i}\lttrans\atr{\xact!j}}},
    \end{math}
    and similarly 
    \begin{math}
      \ell\leq\fcard{\setst{\yact?k}{{\yact?k}\lttrans\atr{\yact!\ell}}}.
    \end{math}
    It suffices to construct an interleaving
    $\btr\in(\btr_1\interleave\btr_2)$ such that whenever $\btr$
    contains a subsequence with names
    \begin{displaymath}
      \xact?j\CC
      \xact!j\CC
      \yact?k\CC
      \yact!k\CC
      \yact?{k+1}\CC
      \yact!{k+1}\CC
      \ldots\CC
      \yact?{k+x}\CC
      \yact!{k+x}
    \end{displaymath}
    then for every $k\leq \ell\leq k+x$, we have
    \begin{displaymath}
      \setst{\xact?i}{{\xact?i}\lttrans\atr{\xact!j}}
      \subseteq
      \setst{\xact?i}{{\xact?i}\lttrans\atr{\yact!\ell}}
    \end{displaymath}
    and symmetrically for subsequences
    \begin{math}
      \yact?k\CC
      \yact!k\CC
      \xact?j\CC
      \xact!j\CC
      \xact?{j+1}\CC
      \xact!{j+1}\CC
      \ldots\CC
      \xact?{j+y}\CC
      \xact!{j+y}.
    \end{math}
    \begin{journal}
    Given such a $\btr$, we know that
    \begin{math}
      j+\ell\leq\fcard{\setst{\xact?i}{{\xact?i}\lttrans\atr{\yact!\ell}}
        \cup\setst{\yact?k}{{\yact?k}\lttrans\atr{\yact!\ell}}},
    \end{math}
    as required. 

    We now demonstrate the existence of such a $\btr$. Define the
    set $\fmerge{\aacts}{\bacts}$ as follows.
    \begin{gather*}
      \begin{aligned}
        \fmerge{\aacts}{\trempty} &= \set{\aacts}\QQQUAD\;\;
        &
        \fmerge{\trempty}{\bacts} &= \set{\bacts}
      \end{aligned}
      \\
      \begin{aligned}
        \fmerge{\aacts\,\xact?j\,\xact!j}{\,\bacts\,\yact?\ell\,\yact!\ell}
        &\ni
        \cacts\,\yact?\ell\,\yact!\ell\;\;
        \begin{array}[t]{l}
          \textif
          \cacts\in\fmerge{\aacts\,\xact?j\,\xact!j}{\,\bacts}
          \\\textand
          \setst{\xact?i}{{\xact?i}\lttrans\atr{\xact!j}}
          \subseteq
          \setst{\xact?i}{{\xact?i}\lttrans\atr{\yact!\ell}}
        \end{array}
        \\
        \fmerge{\aacts\,\xact?j\,\xact!j}{\,\bacts\,\yact?\ell\,\yact!\ell}
        &\ni
        \cacts\,\xact?j\,\xact!j\;\;
        \begin{array}[t]{l}
          \textif
          \cacts\in\fmerge{\aacts}{\,\bacts\,\yact?\ell\,\yact!\ell}
          \\\textand
          \setst{\yact?k}{{\yact?k}\lttrans\atr{\yact!\ell}}
          \subseteq
          \setst{\yact?k}{{\yact?k}\lttrans\atr{\xact!j}}
        \end{array}
      \end{aligned}
    \end{gather*}
    \end{journal}
    To dem\-on\-strate the existence of an appropriate $\btr$, it suffices
    to show that 
    \begin{math}
      \fmerge
      {\xact?1\,\allowbreak\xact!1\,\allowbreak\ldots\,\allowbreak\xact?m\,\allowbreak\xact!m}
      {\allowbreak\yact?1\,\yact!1\,\allowbreak\ldots\,\allowbreak\yact?n\,\allowbreak\yact!n}
    \end{math}
    is nonempty.  By \operationality{}, it must be the case that either
    \begin{enumerate*}
    \item 
      $\xact!j\lttrans\atr\allowbreak\yact!\ell$, in which case
      \begin{math}
        \setst{\xact?i}{{\xact?i}\lttrans\atr{\xact!j}}
        \subseteq
        \setst{\xact?i}{{\xact?i}\lttrans\atr{\yact!\ell}},
      \end{math}
    \item 
      $\yact!\ell\lttrans\atr\xact!j$, in which case
      \begin{math}
        \setst{\yact?k}{{\yact?k}\lttrans\atr{\yact!\ell}}
        \subseteq
        \setst{\yact?k}{{\yact?k}\lttrans\atr{\xact!j}},
      \end{math}
      or
    \item $\xact!j$ and $\yact!\ell$ are unordered, in which case both
      conclusions hold.
    \end{enumerate*}
    Therefore an appropriate $\btr$ exists.
  \end{proofkeep}
\end{theorem}

\begin{journal}
\begin{example}
  We demonstrate the $\fmerge{}{}$ function defined in the proof
  above using the following traces.
  \begin{gather*}
    \begin{aligned}
      \btr_1 &=
      {\BINC1??}
      {\EINC1!0}
      {\BINC2??}
      {\EINC2!1}
      {\BINC3??}
      {\EINC3!2}
      &\btr_2 &=
      {\BINC8??}
      {\EINC8!0}
      {\BINC5??}
      {\EINC5!1}
      {\BINC6??}
      {\EINC6!2}
      \\
      \atr_1 &=
      {\BINC3??}
      {\BINC2??}
      {\EINC2!1}
      {\BINC1??}
      {\EINC1!0}
      {\EINC3!2}
      &\atr_2 &=
      {\BINC6??}
      {\BINC5??}
      {\EINC5!1}
      {\BINC8??}
      {\EINC8!0}
      {\EINC6!2}
    \end{aligned}
    \\
    \atr =
    {\BINC6??}
    {\BINC5??}
    {\EINC5!1}
    {\BINC3??}
    {\BINC8??}
    {\EINC8!0}
    {\BINC2??}
    {\EINC2!1}
    {\BINC1??}
    {\EINC1!0}
    {\EINC3!2}
    {\EINC6!2}
  \end{gather*}
  In the graph below, we draw an edge from $\aact_j$ to $\bact_\ell$
  if
  \begin{math}
    \setst{\xact?i}{{\xact?i}\lttrans\atr{\xact!j}}
    \subseteq
    \setst{\xact?i}{{\xact?i}\lttrans\atr{\yact!\ell}},
  \end{math}
  indicating that $\bact_\ell$ may come after $\aact_j$.
  Edges from $\bact_\ell$ to $\aact_j$ are similar.  When an edge is
  bidirectional, we use a dashed line.
  
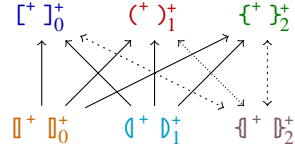
\begin{displaymath}
    \begin{tikzpicture}[xscale=1.5,yscale=-1.5]
      \node (a0) at (0,0) {$\YINC1!0$};
      \node (a1) at (1,0) {$\YINC2!1$};
      \node (a2) at (2,0) {$\YINC3!2$};
      \node (b0) at (0,1) {$\YINC8!0$};
      \node (b1) at (1,1) {$\YINC5!1$};
      \node (b2) at (2,1) {$\YINC6!2$};
      \draw [query] (b0) to (a0);
      \draw [query] (b0) to (a1);
      \draw [query] (b0) to (a2);
      \draw [query] (b1) to (a0);
      \draw [query] (b1) to (a1);
      \draw [query] (b1) to (a2);
      \draw [other] (a2) to (b2); \draw [other] (b2) to (a2);
      \draw [other] (a1) to (b2); \draw [other] (b2) to (a1);
      \draw [other] (a0) to (b2); \draw [other] (b2) to (a0);
    \end{tikzpicture}
  \end{displaymath}
  The following traces are derived from the $\fmerge{}{}$ algorithm.
  \begin{align*}
    {\BINC8??} {\EINC8!0}
    {\BINC5??} {\EINC5!1}
    {\BINC6??} {\EINC6!2}
    {\BINC1??} {\EINC1!0}
    {\BINC2??} {\EINC2!1}
    {\BINC3??} {\EINC3!2}
    \\
    {\BINC8??} {\EINC8!0}
    {\BINC5??} {\EINC5!1}
    {\BINC1??} {\EINC1!0}
    {\BINC6??} {\EINC6!2}
    {\BINC2??} {\EINC2!1}
    {\BINC3??} {\EINC3!2}
    \\
    {\BINC8??} {\EINC8!0}
    {\BINC5??} {\EINC5!1}
    {\BINC1??} {\EINC1!0}
    {\BINC2??} {\EINC2!1}
    {\BINC6??} {\EINC6!2}
    {\BINC3??} {\EINC3!2}
    \\
    {\BINC8??} {\EINC8!0}
    {\BINC5??} {\EINC5!1}
    {\BINC1??} {\EINC1!0}
    {\BINC2??} {\EINC2!1}
    {\BINC3??} {\EINC3!2}
    {\BINC6??} {\EINC6!2}
  \end{align*}
  Suppose instead that we have the following $\atr$.
  \begin{gather*}
    \atr =
    {\BINC6??}
    {\BINC5??}
    {\EINC5!1}
    {\BINC3??}
    {\BINC8??}
    {\BINC2??}
    {\EINC2!1}
    {\EINC8!0}
    {\EINC6!2}
    {\BINC1??}
    {\EINC1!0}
    {\EINC3!2}
  \end{gather*}
  Then the graph and resulting traces are as follows.
  
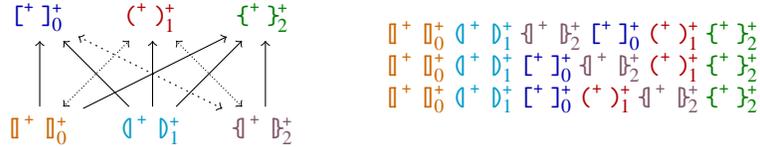
\begin{align*}
    \begin{tikzpicture}[xscale=1.5,yscale=-1.5]
      \node (a0) at (0,0) {$\YINC1!0$};
      \node (a1) at (1,0) {$\YINC2!1$};
      \node (a2) at (2,0) {$\YINC3!2$};
      \node (b0) at (0,1) {$\YINC8!0$};
      \node (b1) at (1,1) {$\YINC5!1$};
      \node (b2) at (2,1) {$\YINC6!2$};
      \draw [query] (b0) to (a0);
      \draw [other] (b0) to (a1); \draw [other] (a1) to (b0);
      \draw [query] (b0) to (a2);
      \draw [query] (b1) to (a0);
      \draw [query] (b1) to (a1);
      \draw [query] (b1) to (a2);
      \draw [query] (b2) to (a2);
      \draw [other] (a1) to (b2); \draw [other] (b2) to (a1);
      \draw [other] (a0) to (b2); \draw [other] (b2) to (a0);
    \end{tikzpicture}
    &&
    \begin{array}{l}
    {\BINC8??} {\EINC8!0}
    {\BINC5??} {\EINC5!1}
    {\BINC6??} {\EINC6!2}
    {\BINC1??} {\EINC1!0}
    {\BINC2??} {\EINC2!1}
    {\BINC3??} {\EINC3!2}
    \\
    {\BINC8??} {\EINC8!0}
    {\BINC5??} {\EINC5!1}
    {\BINC1??} {\EINC1!0}
    {\BINC6??} {\EINC6!2}
    {\BINC2??} {\EINC2!1}
    {\BINC3??} {\EINC3!2}
    \\
    {\BINC8??} {\EINC8!0}
    {\BINC5??} {\EINC5!1}
    {\BINC1??} {\EINC1!0}
    {\BINC2??} {\EINC2!1}
    {\BINC6??} {\EINC6!2}
    {\BINC3??} {\EINC3!2}
    \end{array}
  \end{align*}
  In general, if one where to include the linear order from the
  specification (eg, from ${\YINC1!0}$ to ${\YINC2?1}$), the resulting
  graph might be cyclic, even if the dotted edges were removed.
\end{example}

\subsection{Comparison with \citeauthor{DBLP:conf/popl/HenzingerKPSS13}}
\label{sec:qqc:compare}

QQC does not immediately correspond to any relaxations considered
by \textcite{DBLP:conf/popl/HenzingerKPSS13}.  The comparison is
subtler than it appears at first glance.  The examples in this
subsection are from \textcite{Ali}.

Consider the following two stack traces:
\begin{align*}
  \BPUT3?c
  \YPUT1.a
  \YPUT2.b
  \YGET4.a
  \EPUT3!c
  &&
  \BPUT3?c
  \YPUT1.a
  \YPUT2.b
  \EPUT3!c
  \YGET4.a
\end{align*}
The first of these is QQC with the stack specification
\begin{math}
  \YPUT2.b
  \YPUT1.a
  \YGET4.a
  \BPUT3?c
  \EPUT3!c
\end{math}
whereas the second is not QQC with any stack trace.
In the framework of \parencite{DBLP:conf/popl/HenzingerKPSS13}, these
two traces represent the same relaxed behavior, namely $1$ out-of-order
(when $\COLOR1a$ is popped, at least $\COLOR2b$ must be above
$\COLOR1a$ on the stack).  Thus, QQC makes distinctions that are not
found in \parencite{DBLP:conf/popl/HenzingerKPSS13}.

For stacks, it may be that QQC is finer
than \parencite{DBLP:conf/popl/HenzingerKPSS13}; however, in general
the criteria are unrelated.  In the other direction, consider the
following family of queue traces:
\begin{align*}
  \BPUT3?a
  \YPUT1.{b_1}
  \YPUT1.{b_1}
  \cdots
  \YPUT1.{b_n}
  \YPUT2.c
  \YGET4.c
  \EPUT3!a
\end{align*}
This is QQC with the queue specification
\begin{math}
  \YPUT2.c
  \YPUT1.{b_1}
  \YPUT1.{b_1}
  \cdots
  \YPUT1.{b_n}
  \YGET4.c
  \BPUT3?a
  \EPUT3!a\!\!.
\end{math}
In the framework of \parencite{DBLP:conf/popl/HenzingerKPSS13}, this
would be $n$ out-of-order because at least all $\COLOR1{b_i}$'s should
be in the queue before $\COLOR2c$ is inserted into the queue, so the
removal of $\COLOR2c$ from the queue must happen when there are $n$
elements ahead of $\COLOR2c$ in the queue.
Thus, \parencite{DBLP:conf/popl/HenzingerKPSS13} makes distinctions
that are not found in QQC.

\section{Stack example}
\label{sec:stack}

We show that, under reasonable assumptions, our \fstack{N} is QQC.  We
extend this argument to the elimination-tree stacks
of \parencite{DBLP:journals/mst/ShavitT97}.

In proving that executions of our \fstack{N} are QQC, the key step is
to generate the corresponding specification trace.  To do so, we
consider the following instrumentation.
\begin{lstlisting}[numbers=left]
class #\ctype{Stack<N:Int>}# {
  field b:#\ctype{[0..N-1]}# = 0;                 // 1 balancer
  field s:#\ctype{Stack[]}#  = [[], [], ..., []]; // N stacks of values
  field e:#\ctype{[0..N-1]}# = 0;                 // 1 emitter
  field q:#\ctype{Queue[]}#  = [[], [], ..., []]; // N queues of actions
  method push(x:Object):Unit {
    val i:#\ctype{[0..N-1]}#;
    atomic {i=b; b++;}#\label{a1a}#
    atomic {val v=s[i].push(x); q[i].add("push" x); emit(); return v;} }#\label{a2a}#
  method pop():Object {
    val i:#\ctype{[0..N-1]}#;
    atomic {i=b-1; b--;}#\label{a1b}#
    atomic {val v=s[i].pop(); q[i].add("pop" v); emit(); return v;} }#\label{a2b}#             
  method emit():Unit {
    while (q[e].first()=~"push" || q[e-1].first()=~"pop") {
      if (q[e].first()=~"push")  {print (q[e].remove());   e++;}
      if (q[e-1].first()=~"pop") {print (q[e-1].remove()); e--;} } } }
\end{lstlisting}
The state of the machine includes the values of the balancer
\texttt{b} and stacks \texttt{s}.  It also includes queues \texttt{q}
to store the actions that have been executed on each stack and a
\emph{emitter} \texttt{e}, with the same range as \texttt{b}, which
indicates the queue that should produce the next specification
action. The emitter prints any completed pushes from \texttt{s[e]} and
any completed pops from \texttt{s[e-1]}.  When the emitter prints a
push, it removes it from the queue and increments \texttt{e}; when it
prints a pop, it removes it from the queue and decrements \texttt{e}.
Emitter actions take place as soon as possible, and the emitter
continues until it has nothing left to do.

Atomic blocks can only execute concurrently if they do not touch the
same shared state.  For the code in the introduction, this imposes an
order between all executions of the first atomic (lines \ref{a1a} and
\ref{a1b}), since they touch the shared variable \texttt{b}; order is
only imposed between executions of the second atomic that update the
same stack. The presence of \texttt{emit} indicates also imposes an
order between all executions of the second atomic (lines
\ref{a2a} and \ref{a2b}), since \texttt{emit} touches the shared
variable \texttt{e}.  This total order on calls to \texttt{emit}
ensures that the printed trace is indeed a stack trace, as we argue below.

\begin{definition}
  Let $\aact$ be a call to \texttt{push} or \texttt{pop}. Then
  $\ftime1\aact$ is the time of the execution of the first atomic
  statement in the \fstack{N}, and $\ftime2\aact$ is the time of the
  execution of the second atomic. A \emph{linearized trace} of an
  \fstack{N} is one in which the invocations are ordered consistently
  with $\ftime1{}$ and the responses are ordered consistently with
  $\ftime2{}$.
\end{definition}

For example, from the linearization
\begin{math}
  \BPUT2?b \YPUT1.a \EPUT2!b
\end{math}
we know
\begin{math}
  {\ftime1{\BPUT2?b}} < {\ftime1{\BPUT1?a}}
\end{math}
and
\begin{math}
  {\ftime2{\EPUT1!a}} < {\ftime2{\EPUT2!b}}.
\end{math}
Such a linearized trace is distinct from other linearizations of the
same trace, such as
\begin{math}
  \BPUT2!b \BPUT1?a \EPUT2!b \EPUT1?a\text{,}\;\allowbreak
  \BPUT1?a \BPUT2!b \EPUT1?a \EPUT2!b\textand
  \BPUT1?a \YPUT2.b \EPUT1!a.
\end{math}

The response order in the linearized trace is particularly
significant.  For example, the linearization
\begin{math}
  \BPUT2?b \YPUT1.a \EPUT2!b \YGET1.a \YGET2.b
\end{math}
cannot result from the execution of a \fstack{1}.  In this case
$\COLOR1a$ is pushed before $\COLOR2b$ and therefore the pop of
$\COLOR1a$ cannot be ordered before the pop of $\COLOR2b$.

\begin{example}
  Consider the following linearized trace of a \fstack{2}.
  \begin{displaymath}
    \BPUT2?c
    \BPUT4?b
    \EPUT4!b
    \BPUT1?a
    \EPUT1!a
    \EPUT2!c
    \BGET2?c
    \EGET2!c
    \BGET4?b
    \EGET4!b
    \BGET1?a
    \EGET1!a
  \end{displaymath}
  Execution proceeds as follows.  We show the atomic that is being
  executed above the arrow.  Arrows without labels are executed within
  \texttt{emit}, atomically with the prior label.  On the right-hand
  side, we show any emitted actions, followed by the resulting state.
  The initial state of the machine is $\sstate{0}{}{}{}{}{0}$.
  \begin{alignat*}{2}
    &&&\sxstate{0}{}{}{}{}{0}
    \\\xrightarrow{\BPUT2?c}\;\; &         &&\sxstate{1}{}{}{}{}{0}
    \\\xrightarrow{\BPUT4?b}\;\; &         &&\sxstate{0}{}{}{}{}{0}
    \\\xrightarrow{\EPUT4!b}\;\; &         &&\sxstate{0}{}{\COLOR4b}{}{\YPUT4?b}{0}
    \\\xrightarrow{\BPUT1?a}\;\; &         &&\sxstate{1}{}{\COLOR4b}{}{\YPUT4?b}{0}
    \\\xrightarrow{\EPUT1!a}\;\; &         &&\sxstate{1}{\COLOR1a}{\COLOR4b}{\YPUT1?a}{\YPUT4?b}{0}
    \\\xrightarrow{}        \;\; &\YPUT1?a &&\sxstate{1}{\COLOR1a}{\COLOR4b}{}{\YPUT4?b}{1} 
    \\\xrightarrow{}        \;\; &\YPUT4?b &&\sxstate{1}{\COLOR1a}{\COLOR4b}{}{}{0}
    \\\xrightarrow{\EPUT2!c}\;\; &         &&\sxstate{1}{\COLOR2c\COLOR1a}{\COLOR4b}{\YPUT2?c}{}{0}
    \\\xrightarrow{}        \;\; &\YPUT2?c &&\sxstate{1}{\COLOR2c\COLOR1a}{\COLOR4b}{}{}{1}
    \\\xrightarrow{\BGET2?c}\;\; &         &&\sxstate{0}{\COLOR2c\COLOR1a}{\COLOR4b}{}{}{1}
    \\\xrightarrow{\EGET2!c}\;\; &         &&\sxstate{0}{\COLOR1a}{\COLOR4b}{\YGET2?c}{}{1}
    \\\xrightarrow{}        \;\; &\YGET2?c &&\sxstate{0}{\COLOR1a}{\COLOR4b}{}{}{0}
    \\\xrightarrow{\BGET4?b}\;\; &         &&\sxstate{1}{\COLOR1a}{\COLOR4b}{}{}{0}
    \\\xrightarrow{\EGET4!b}\;\; &         &&\sxstate{1}{\COLOR1a}{}{}{\YGET4?b}{0}
    \\\xrightarrow{}        \;\; &\YGET4?b &&\sxstate{1}{\COLOR1a}{}{}{}{1}
    \\\xrightarrow{\BGET1?a}\;\; &         &&\sxstate{0}{\COLOR1a}{}{}{}{1}
    \\\xrightarrow{\EGET1!a}\;\; &         &&\sxstate{0}{}{}{\YGET1?a}{}{1}
    \\\xrightarrow{}        \;\; &\YGET1?a &&\sxstate{0}{}{}{}{}{0} 
    \amsqed
  \end{alignat*}
\end{example}
\begin{example}
  Consider the following execution of the instrumented counter.
  \begin{alignat*}{2}
    &&&\systate{0}{}{}{}{}{0}
    \\\xrightarrow{\BPUT1?0}\;\; &         &&\systate{1}{}{}{}{}{0}
    \\\xrightarrow{\BPUT2?a}\;\; &         &&\systate{0}{}{}{}{\YPUT2?a}{0}
    \\\xrightarrow{\EPUT2!a}\;\; &         &&\systate{0}{}{\COLOR2a}{}{\YPUT2?a}{0}
    \\\xrightarrow{\BGET2?a}\;\; &         &&\systate{1}{}{\COLOR2a}{}{\YPUT2?a}{0}
    \\\xrightarrow{\EGET2?a}\;\; &         &&\systate{1}{}{}{}{\YPUT2?a\YGET2?a}{0}
    \\\xrightarrow{\BPUT2?b}\;\; &         &&\systate{0}{}{}{}{\YPUT2?a\YGET2?a}{0}
    \\\xrightarrow{\EPUT2?b}\;\; &         &&\systate{0}{}{\COLOR2b}{}{\YPUT2?a\YGET2?a\YPUT2?b}{0}
    \\\xrightarrow{\BPUT1?2}\;\; &         &&\systate{1}{}{\COLOR2b}{}{\YPUT2?a\YGET2?a\YPUT2?b}{0}
    \\\xrightarrow{\BPUT2?c}\;\; &         &&\systate{0}{}{\COLOR2b}{}{\YPUT2?a\YGET2?a\YPUT2?b}{0}
    \\\xrightarrow{\EPUT2?c}\;\; &         &&\systate{0}{}{\COLOR2b\COLOR2c}{}{\YPUT2?a\YGET2?a\YPUT2?b\YPUT2?c}{0}
    \\\xrightarrow{\EPUT1?0}\;\; &         &&\systate{0}{\COLOR10}{\COLOR2b\COLOR2c}{\YPUT1?0}{\YPUT2?a\YGET2?a\YPUT2?b\YPUT2?c}{0}
    \\\xrightarrow{}        \;\; &\YPUT1?0 &&\systate{0}{\COLOR10}{\COLOR2b\COLOR2c}{}{\YPUT2?a\YGET2?a\YPUT2?b\YPUT2?c}{1}
    \\\xrightarrow{}        \;\; &\YPUT2?a &&\systate{0}{\COLOR10}{\COLOR2b\COLOR2c}{}{\YGET2?a\YPUT2?b\YPUT2?c}{0}
    \\\xrightarrow{}        \;\; &\YGET2?a &&\systate{0}{\COLOR10}{\COLOR2b\COLOR2c}{}{\YPUT2?b\YPUT2?c}{1}
    \\\xrightarrow{}        \;\; &\YPUT2?b &&\systate{0}{\COLOR10}{\COLOR2b\COLOR2c}{}{\YPUT2?c}{0}
    \\\xrightarrow{\EPUT1?1}\;\; &         &&\systate{0}{\COLOR10\COLOR11}{\COLOR2b\COLOR2c}{\YPUT1?1}{\YPUT2?c}{0}
    \\\xrightarrow{}        \;\; &\YPUT1?1 &&\systate{0}{\COLOR10\COLOR11}{\COLOR2b\COLOR2c}{}{\YPUT2?c}{1}
    \\\xrightarrow{}        \;\; &\YPUT2?c &&\systate{0}{\COLOR10\COLOR11}{\COLOR2b\COLOR2c}{}{}{0}
  \end{alignat*}
  This produces the following linearized trace $\atr$ and
  specification $\btr$.
  \begin{align*}
    \atr&=\BPUT1?0 \YPUT2?a \YGET2?a \YPUT2?b \BPUT1?1 \YPUT2?c \EPUT1?0 \EPUT1?1
    \\
    \btr&=\YPUT1?0 \YPUT2?a \YGET2?a \YPUT2?b \YPUT1?1 \YPUT2?c
  \end{align*}
  After the push of $\COLOR2c$ returns, we have
  \begin{math}
    \texttt{q[1]}= [\YPUT2?a\YGET2?a\YPUT2?b\YPUT2?c].
  \end{math}
  When the first $\EPUT1!0$ occurs, the first three actions in the
  \texttt{q[1]} must be emitted.
\end{example}

\begin{lemma}
  \label{thm:qqc:instr:qqc}
  Given an instrumented execution of an \fstack{N}, the linearized
  trace of the execution is QQC with the emitted specification.
  \begin{proofsketch}
    Let us refer to a sequence like 
    \begin{math}
      \YPUT2?a\YGET2?a\YPUT2?b
    \end{math}
    as a \emph{chain}.  A chain is a sequence of calls that can be
    emitted from a single queue without any intervening change to
    \texttt{e}.  By \autoref{thm:qqc:2} suffices to show that after
    the execution of each atomic, the number of chains is bounded by
    the number of open calls.  This follows by induction on the length
    of the instrumented execution.    
  \end{proofsketch}
\end{lemma}

In light of \autoref{thm:qqc:instr:qqc}, to show that the \fstack{N} is
QQC, it suffices to show that the emitted specification is indeed a
stack specification.  Unfortunately, as observed
in \parencite{DBLP:journals/mst/ShavitT97}, this fails to hold.
\begin{example}
  \label{ex:2stack:notqqc}
  As discussed in \autoref{ex:2stack:notqqc1},
  the linearized trace
  \begin{math}
    \YPUT1.a
    \YPUT2.b
    \BPUT1?c
    \YGET1.a
    \EPUT1!c
  \end{math}
  generates the specification
  \begin{math}
    \YPUT1.a
    \YPUT2.b
    \YGET1.a
    \BPUT1?c
    \EPUT1!c.
  \end{math}
  However, this specification is not a stack trace.
  With some number of initial pushes, this execution is still
  possible:  The linearized trace
  \begin{math}
    \YPUT1.x
    \YPUT2.y
    \YPUT1.a
    \YPUT2.b
    \BPUT1?c
    \YGET1.a
    \EPUT1!c
  \end{math}
  generates the specification
  \begin{math}
    \YPUT1.x
    \YPUT2.y
    \YPUT1.a
    \YPUT2.b
    \YGET1.a
    \BPUT1?c
    \EPUT1!c.
  \end{math}
\end{example}

This problematic execution occurs because a push and pop are racing at
the first stack, yet the pop retrieves a prior value: the pop has
\emph{overtaken} the push.  
We must disallow such
executions. It is not sufficient to require only that pop operations
block on an empty stack.

\begin{definition}
  An execution is \emph{properly-popped} if for every push $\aact$ and
  pop $\bact$ that are assigned the same stack \texttt{s[i]},
  \begin{displaymath}
    \ftime1\aact<\ftime1\bact \textimplies
    \ftime2\aact<\ftime2\bact.
  \end{displaymath}
\end{definition}

\begin{lemma}
  \label{thm:qqc:instr:stack}
  If an execution of the instrumented \fstack{N} is properly-popped,
  then it trace it prints is a stack trace.
  \begin{proofsketch}
    It is sufficient to note that the execution of the emitter follows
    the same pattern as the uninstrumented \fstack{N} on a sequential
    execution.  (This is only true with proper popping.)  The result
    follows since, as shown
    in \parencite{DBLP:journals/mst/ShavitT97}, the sequential
    execution of the \fstack{N} does simulate a stack.
  \end{proofsketch}
\end{lemma}

\begin{theorem}
  \label{thm:nstack}
  Any properly-popped execution of an \fstack{N} is QQC.
  \begin{proof}
    By \autorefs{thm:qqc:instr:qqc} and \ref{thm:qqc:instr:stack}.
  \end{proof}
\end{theorem}

We have shown that for properly-popped executions (where a pop may not
ignore a concurrent push on the same stack) the \fstack{N} is QQC.
As noted in the introduction, we know of no analogous
condition for increment/decrement counters.

In \parencite{DBLP:journals/mst/ShavitT97}, \citeauthor{DBLP:journals/mst/ShavitT97} show
that in a quiescent state, their elimination-tree stack reaches a
state consistent with a stack.  We now consider the relation between
our \fstack{N}s and these elimination-tree stacks.
\begin{example}
A depth-2 elimination-tree stack can be implemented using three
a\-tom\-ic booleans---top (\texttt{t}), left (\texttt{l}) and right (\texttt{r})---and 4
linearizable stacks with addresses \sW{}, \sX{}, \sY{} and \sZ{}.  
\begin{displaymath}
  \begin{tikzpicture}[font=\ttfamily,thick,level/.style={sibling distance=50mm/#1}]
    \node[vertex] {t}
    child { node[vertex] {l} 
      child { node[vertex] {00} }
      child { node[vertex] {01} }
    }
    child { node[vertex] {r} 
      child { node[vertex] {00} }
      child { node[vertex] {01} }
    }
    ;
  \end{tikzpicture}
\end{displaymath}
The \emph{address} of a stack in an depth-$d$ elimination tree is a
sequence of $d$ booleans, indicating the value of the boolean at each
level, going down a branch of the tree.
Both \texttt{push} and \texttt{pop} toggle the booleans as they go
down the tree, using an atomic read and update. If $\texttt{t}=0$,
then \texttt{push} sets $\texttt{t}=1$ and goes {left}.  If
$\texttt{t}=0$, then \texttt{pop} sets $\texttt{t}=1$ and goes
{right}.  The methods follow this same pattern down the tree
until they reach the bottom-level stack, at which point they perform
the operation.  Initially all booleans are set to $0$.
For example, one uninstrumented execution proceeds as follows.
\begin{alignat*}{3}
                           &\ststate{0}{0}{}{}{0}{}{}
  \\\xrightarrow{\BPUT2?e} &\ststate{1}{0}{}{}{0}{}{}
  \\\xrightarrow{\YPUT1.b} &\ststate{0}{0}{}{}{1}{\COLOR1b}{}
  \\\xrightarrow{\YPUT1.a} &\ststate{1}{1}{\COLOR1a}{}{1}{\COLOR1b}{}
  \\\xrightarrow{\YPUT1.d} &\ststate{0}{1}{\COLOR1a}{}{0}{\COLOR1b}{\COLOR1d}
  \\\xrightarrow{\YPUT1.c} &\ststate{1}{0}{\COLOR1a}{\COLOR1c}{0}{\COLOR1b}{\COLOR1d}
  \\\xrightarrow{\EPUT2?e} &\ststate{1}{1}{\COLOR2e\COLOR1a}{\COLOR1c}{0}{\COLOR1b}{\COLOR1d}
  \\\xrightarrow{\YGET3?e} &\ststate{0}{0}{\COLOR1a}{\COLOR1c}{0}{\COLOR1b}{\COLOR1d}
  \\\xrightarrow{\YGET3?d} &\ststate{1}{0}{\COLOR1a}{\COLOR1c}{1}{\COLOR1b}{}
  \\\xrightarrow{\YGET3?c} &\ststate{0}{1}{\COLOR1a}{}{1}{\COLOR1b}{}
  \\\xrightarrow{\YGET3?b} &\ststate{1}{1}{\COLOR1a}{}{0}{}{}
  \\\xrightarrow{\YGET3?a} &\ststate{0}{0}{}{}{0}{}{}
\end{alignat*}
This gives the trace
\begin{math}
  \BPUT2?e
  \YPUT1.b
  \YPUT1.a
  \YPUT1.d
  \YPUT1.c
  \EPUT2!e
  \YGET3.e
  \YGET3.d
  \YGET3.c
  \YGET3.b
  \YGET3.a
\end{math}
which is QQC with respect to 
\begin{math}
  \YPUT1.a
  \YPUT1.b
  \YPUT1.c
  \YPUT1.d
  \YPUT2.e
  \YGET3.e
  \YGET3.d
  \YGET3.c
  \YGET3.b
  \YGET3.a.
\end{math}
Our \fstack{4} does not generate this execution trace; however, our
\fstack{2} does.  In general, our \fstack{N^d} has strictly fewer
behaviors than the $N$-branching elimination-tree stack of depth $d$.
We leave open the question of whether a $N$-branching elimination-tree
stack of depth $d$ has behaviors that not possible for an \fstack{N}.
\end{example}

The instrumented execution of a $N$-branching elimination-tree stack
of depth ${d>1}$ can be defined using the execution of
elimination-tree stacks of depth $d-1$, using the same strategy as our
$N$-\texttt{Stack}.  While the balancer's behavior is more general in
the composed system, the emitter's is not: The emitter code is
entirely sequentialized, therefore a $2$-nested $N$-branching emitter has
the same behavior as a flat $N^2$-branching emitter.
\begin{theorem}
  Any properly-popped execution of a $N$-branching elimination-tree
  stack of depth $d$ is QQC.  
  \begin{proofsketch}
    Following the strategy in \autoref{thm:nstack}, we need only prove
    the corresponding lemmas.  In each case, the proof procedes by
    induction on $d$.  In each case the basis is the same: 
    a depth $1$ elimination tree stack is simply an $N$-\texttt{Stack}.

    The analogue of \autoref{thm:qqc:instr:qqc} follows, as
    before, by induction on the length of the instrumented execution.
    An open call at depth $d$ may initiate a new chain, but only in
    \emph{one} stack of depth $d-1$.
    
    For the analogue of \ref{thm:qqc:instr:stack} it suffices to
    observe that the emitter's behavior is the same if levels $d>1$
    and $d-1$ are flattened into a single level of size $N^2$.  This
    follows from the atomicity of the emitter.
  \end{proofsketch}
\end{theorem}
\end{journal}

\section{Conclusions}
\label{sec:end}

\emph{Quantitative quiescent consistency (QQC)}  is a correctness criterion
for concurrent data structures that relaxes linearizability and
refines quiescent consistency.  To the best of our knowledge, it is
the first such criterion to be proposed.

To show that QQC is a robust concept, we have provided three alternate
characterizations:
\begin{enumerate*}
\item in the style of linearizability,
\item counting the number of calls before a return, and
\item using speculative flat combining.
\end{enumerate*}
We have also proven
compositionality (in the style of
\textcite{DBLP:journals/toplas/HerlihyW90}) and\ifconference{, in the full paper,} the correctness 
of data structures defined
by \textcite{DBLP:journals/jacm/AspnesHS94}
and \textcite{DBLP:journals/mst/ShavitT97}.

In order to establish the correctness of the elimination-tree stack
of \parencite{DBLP:journals/mst/ShavitT97}, we had to restrict
attention to traces in which no pop \emph{overtakes} a push on the
same stack.
\ifconference{(The formalities are given in the full paper.)}
A related constraint appears
in a footnote of \parencite{Shavit:2011:DSM:1897852.1897873}:
``To keep things simple, pop operations should block until a matching
push appears.'' 
This, however, is not strong enough to guarantee quiescent
consistency as we have defined it.
Our analysis provides a full account: The stack
is QQC with the no-overtaking requirement and only weakly quiescently consistent without it.

There are many unanswered questions, chief among them: Is QQC useful
in reasoning about client programs?  Is there a verification
methodology for QQC analogous to that developed for linearizability?
Are there other useful data structures that can be shown to satisfy
QQC?

\ifjournal{Linearizability has proven to be a valuable foundation for program
verification techniques.  It remains to be seen if QQC can be of use
in this regard.}

Linearizability is, at its core, \emph{linear}.  We have defined QQC
in terms of general partial orders, and yet the results reported here
are stated in terms of sequential specifications.  Partly we have done
this so that we can relate the definition of QQC to the vast amount of
existing work on linearizability.  However, the general case is
interesting.

\ifjournalelse{\section*}{\paragraph*}{Acknowledgements.}
Gustavo Petri participated in the early discussions motivating
this work.  Alexey Gotsman suggested the connection to flat
combining.  Ali Sezgin provided a comparison
with \parencite{DBLP:conf/popl/HenzingerKPSS13}. 
\ifjournal{Mike Dodds pointed out the connection with \parencite{BDG13}.}
We also thank Alan
Jeffrey, Corin Pitcher and Hongseok Yang for useful discussion.


\printbibliography
\end{document}
